\documentclass[conference]{IEEEtran}
\IEEEoverridecommandlockouts
\usepackage{cite}
\usepackage{amsmath,amssymb,amsfonts}
\usepackage{algorithm}
\usepackage[noend]{algpseudocode}
\usepackage{graphicx}
\usepackage{balance}
\usepackage{xcolor}
\def\BibTeX{{\rm B\kern-.05em{\sc i\kern-.025em b}\kern-.08em
    T\kern-.1667em\lower.7ex\hbox{E}\kern-.125emX}}

\usepackage[utf8]{inputenc}
\usepackage[small]{caption}
\usepackage{booktabs}
\usepackage{url}

\newcommand{\model}{\textsc{Pieck}}
\newcommand{\modelI}{\textsc{PieckIpe}}
\newcommand{\modelII}{\textsc{PieckUea}}
\newcommand{\FRA}{\textsc{FedRecA}}
\newcommand{\PIP}{\textsc{PipA}}
\newcommand{\AHUM}{\textsc{A-hum}}
\newcommand{\ARA}{\textsc{A-ra}}

\newcommand{\NODEF}{\textsc{NoDefense}}
\newcommand{\NB}{\textsc{NormBound}}
\newcommand{\MEDIAN}{\textsc{Median}}
\newcommand{\TMEAN}{\textsc{TrimmedMean}}
\newcommand{\KRUM}{\textsc{Krum}}
\newcommand{\MKRUM}{\textsc{MultiKrum}}
\newcommand{\BULYAN}{\textsc{Bulyan}}
\usepackage{caption} 
\usepackage{multirow} 
\usepackage[normalem]{ulem}
\useunder{\uline}{\ul}{}
\usepackage{subfigure}
\usepackage{float}
\usepackage{bbding} 
\usepackage{colortbl} 
\usepackage{hyperref} 
\hypersetup{
    hidelinks,
}
\usepackage[capitalize]{cleveref}
\usepackage[shortlabels]{enumitem}
\setenumerate[1]{itemsep=0pt,partopsep=0pt,parsep=\parskip,topsep=0.5pt}
\setitemize[1]{itemsep=0pt,partopsep=0pt,parsep=\parskip,topsep=0.5pt}
\setdescription{itemsep=0pt,partopsep=0pt,parsep=\parskip,topsep=0.5pt}

\let\temp\rmdefault
\usepackage{mathpazo}
\let\rmdefault\temp
\newtheorem{property}{\bf Property}

\begin{document}

\title{
Preventing the Popular Item Embedding Based Attack in Federated Recommendations
}
\author{
Jun Zhang$^{\dagger,\ddagger}$ \hspace{1em} Huan Li$^{\dagger,\ddagger,*}$  \hspace{1em} Dazhong Rong$^{\ddagger}$  \hspace{1em} Yan Zhao$^{\S}$ \hspace{1em} Ke Chen$^{\dagger,\ddagger}$ \hspace{1em} Lidan Shou$^{\dagger,\ddagger,*}$ \thanks{*Huan Li and Lidan Shou are the corresponding authors.}
 \vspace{1mm}\\
\fontsize{10}{10}\selectfont $^\dagger$\textit{The State Key Laboratory of Blockchain and Data Security, Zhejiang University}, Hangzhou, China\\
\fontsize{10}{10}\selectfont $^\ddagger$\textit{College of Computer Science and Technology, Zhejiang University}, Hangzhou, China\\
\fontsize{10}{10}\selectfont $^\S$\textit{Department of Computer Science, Aalborg University}, Aalborg, Denmark\\
\fontsize{8}{8}\selectfont\ttfamily\upshape \{zj.cs, lihuan.cs, rdz98, 	chenk, should\}@zju.edu.cn, yanz@cs.aau.dk
}

\maketitle

\thispagestyle{plain}
\pagestyle{plain}

\begin{abstract}
Privacy concerns have led to the rise of federated recommender systems (FRS), which can create personalized models across distributed clients. However, FRS is vulnerable to poisoning attacks, where malicious users manipulate gradients to promote their target items intentionally. 
Existing attacks against FRS have limitations, as they depend on specific models and prior knowledge, restricting their real-world applicability.
In our exploration of practical FRS vulnerabilities, we devise a model-agnostic and prior-knowledge-free attack, named \model{} (Popular Item Embedding based Attack). 
The core module of \model{} is popular item mining, which leverages embedding changes during FRS training to effectively identify the popular items.
Built upon the core module, \model{} branches into two diverse solutions:
The \modelI{} solution employs an item popularity enhancement module, which aligns the embeddings of targeted items with the mined popular items to increase item exposure. 
The \modelII{} further enhances the robustness of the attack by using a user embedding approximation module, which approximates private user embeddings using mined popular items.
Upon identifying \model{}, we evaluate existing federated defense methods and find them ineffective against \model{}, as poisonous gradients inevitably overwhelm the cold target items. 
We then propose a novel defense method by introducing two regularization terms during user training, which constrain item popularity enhancement and user embedding approximation while preserving FRS performance.
We evaluate \model{} and its defense across two base models, three real datasets, four top-tier attacks, and six general defense methods, affirming the efficacy of both \model{} and its defense.
\end{abstract}


\section{Introduction}
\label{sec:intro}

Growing concerns about privacy protection have sparked the advent of federated recommendation as a novel paradigm for creating personalized recommender models across distributed clients \cite{MintoHLH21,FedRec++,WangYCYZZ22,FedFast}.
In Federal Recommendation Systems (FRS), the model is bifurcated: the non-shareable client component, housing the user's embedding\footnote{A term used synonymously with `embedding vectors' in this paper.}, a.k.a. sensitive personalized data; and the shareable component interchanges the public items' embeddings with the server, along with an \emph{interaction function} designed to predict user-item pairs.
Nevertheless, FRS remains susceptible to \textbf{targeted poisoning attacks} \cite{fedrecattack,pipattack,a-hum}, which use malicious clients to poison the shared model by uploading manipulated gradients. These attacks aim to boost the exposure of selected items in users' recommendation lists. Although these tactics may seem negligible due to the minor impact on model performance, they inflict more harm on users than those untargeted attacks (only aiming for performance degradation) \cite{FL-Defender,CONTRA,FungYB20}.

\begin{figure}[!htbp]
\centering
\includegraphics[width=\linewidth]{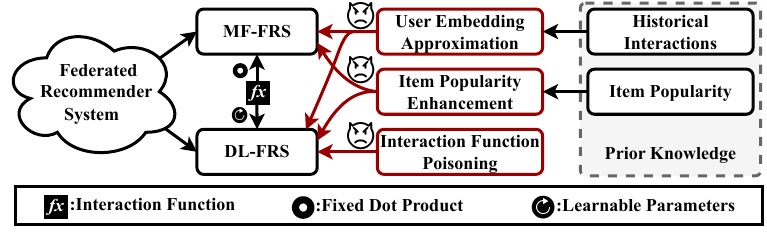}
    \caption{Targeted model poisoning attacks against FRS.}
    \label{fig:attacker}
\end{figure}

\cref{fig:attacker} provides an overall picture of targeted poisoning attacks. 
Two main types of FRS exist: matrix factorization-based FRS (MF-FRS) and deep learning-based FRS (DL-FRS).
Interaction function poisoning~\cite{a-hum} based attacks exclusively target DL-FRS with an assumption that a learnable interaction function exists\footnote{As to be presented in \cref{ssec:preliminaries}, MF-FRS do not use learnable functions but instead a fixed dot product.}. While the other attacks based on user embedding approximation~\cite{fedrecattack} or item popularity enhancement~\cite{pipattack} are model-agnostic, they assume prior knowledge of historical user-item interaction data or item popularity levels. These assumptions, practically inapplicable (no sense to provide such information to attackers), do not lend the corresponding attacks to real-world applications. 
We defer detailed discussions on these attacks to \cref{sec:related}.
This prompts our exploration into \emph{whether a model-agnostic, prior-knowledge-free attack could pose potential risks to realistic FRS and if a proactive defense method could be implemented to combat such threats?}

The development of a model-agnostic attack requires generating poisonous gradients related to item embeddings rather than the interaction function, as the latter is only applicable to DL-FRS. Additionally, the attack should mine knowledge from within the federation due to the unavailability of prior knowledge for poisoning. 
Keeping these points in mind, we identify an effective approach to mining popular items as usable information during FRS training. The mined popular items prove effective in deriving poisonous item embedding gradients, culminating in our attack solution, \model{} (Popular Item Embedding based attaCK).
In particular, \model{}'s core module, named \textbf{popular item mining} (cf.\ \cref{ssec:popular_items}), is driven by a unique characteristic we found --- popular items typically have more significant and long-lasting changes in their embeddings during FRS training.
The mined popular items renovate existing attack methods in a piror-knowledge-free paradigm.
We first propose the \textbf{item popularity enhancement} module (cf.\ \cref{ssec:popular_enhancement}), which aligns the embeddings of target items with those of mined popular items, thus increasing the exposure of target items. The resulting solution is coined \modelI{}.
However, \modelI{} may degrade while the FRS converges towards prominent personalization for each client.
We then propose \modelII{}, which substitutes the item popularity enhancement module with a \textbf{user embedding approximation} module (cf.\ \cref{ssec:user_embed_app}). This module approximates private user embeddings with mined popular item embeddings to derive poisonous item gradients, reflecting our discovery that popular items and most users share a close embedding distribution within the symmetric FRS model.
Notedly, \modelII{} can achieve higher attack effectiveness than \modelI{}. 
However, \modelI{} fewer resources in computations, per our cost analysis in \cref{ssec:cost_analysis}.

In response to the identified attack, we evaluate the efficacy of existing defense methods~\cite{NormBound,Media-TrimmedMean,Krum-MultiKrum,Bulyan} in federated learning. We identify these defenses' limitations, as they require normal gradients to outnumber poisonous ones for a specific item --- an ineffective strategy against \model{}, where poisonous gradients commonly overwhelm a target item. An empirical validation is provided in \cref{exp:defense_performance}.
Following this, we propose a new design, introducing two regularization terms into normal client training to counteract potentially poisonous gradients from the item popularity enhancement and user embedding approximation. We guarantee that this method preserves recommendation performance by incorporating the original training loss.
Extensive experiments across MF-FRS and DL-FRS, three real-world datasets, four top-tier attacks, and six general defense methods, affirm the effectiveness and efficiency of both \model{} and its defense.

In summary, our primary contributions are:
\begin{itemize}[leftmargin=*]
\item We propose \model{}, a model-agnostic and prior-knowledge-free attack against FRS. \model{} lies on automatically mining popular items during training, and branches to two diverse versions that rely on item popularity enhancement and user embedding approximation, respectively (\cref{sec:attack}).
\item We examine the inadequacy of existing federated defense methods via a theoretical analysis, and design a new defense method that is specifically designed for hazardous model poisoning attacks in FRS (\cref{sec:defense}).
\item We conduct extensive experiments across different models, datasets, attack and defense methods, demonstrating the effectiveness and efficiency of our proposals (\cref{sec:experiments}).
\end{itemize}

\section{Related Work} \label{sec:related}

\noindent\textbf{Federated Recommendations}.
In the past decades, recommender systems~\cite{NCF,NGCF,LightGCN,VBPR,CKE,MMGCN} have found widespread applications in facilitating people's discovery of desired information in various domains, ranging from movies on Netflix~\cite{netflix} and videos on YouTube~\cite{youtube}, to products on Taobao~\cite{taobao}.
Typically, a recommender system collects users' historical interactions to train a model, which is subsequently utilized to predict scores, usually ranging from $[0,1]$, for items that a user has not yet interacted with, indicating the user's preferences for those corresponding items.

Motivated by the growing emphasis on user privacy, \emph{federated recommender systems} (FRS) have gained significantattention~\cite{FedGNN,FedRec,FedRec++,ammad2019federated,FedFast,Wu0HNWCYZ21,MintoHLH21}.
Specifically, an FRS splits the model into two parts, namely the client-resident \textbf{personalized model} that retains the sensitive user information and historical interactions, and the \textbf{global model} that exchanges the public item information and gradients of publicly available parameters with the federation via a server.

Two main types of FRS are matrix factorization-based FRS (MF-FRS)~\cite{ammad2019federated,FedRec,FedRec++,9162459} and deep learning-based FRS (DL-FRS)~\cite{WangYCYZZ22,perifanis2022federated,jiang2022fedncf}.
In MF-FRS, users and items are represented as embedding vectors (or simply, embeddings).
The score between a user and an item is computed using a fixed \emph{interaction function} (e.g., dot product) based on their embeddings. 
The global and personalized models encompass separate embeddings for items and users.
DL-FRS follows a similar approach to handle item and user embeddings.
However, in DL-FRS, the interaction function is learnable through deep neural networks. 
Consequently, its global model additionally encompasses learnable parameters of the interaction function, enabling collaborative training with other clients.

\smallskip
\noindent\textbf{Attack Methods against FRS}. 
Many attacks~\cite{Li16Data,HuangMGL0X21,Fang18Po,Fang20In,kapoor2017review} have been proposed against conventional recommender systems.
These attacks involve manipulating user behavior to interact with specific items, thereby poisoning the interaction data.
When a recommender model is trained on poisonous data, it tends to assign abnormally high scores to the target items, leading to their exposure to many benign users.
However, in FRS where gradients are shared instead of users' raw data, most of these attacks~\cite{Li16Data,HuangMGL0X21,Fang18Po,Fang20In} are inapplicable as they rely on access to historical interactions of benign users.
A few studies including~\cite{kapoor2017review} do not require such access but offer very limited effectiveness.
All in all, conventional attacks based on poisonous data fall short in FRS.

Nevertheless, FRS encounters the peril of \textbf{model poisoning attacks}, which manipulate malicious users to upload poisonous gradients to corrupt the recommender model.
Model poisoning attacks are further classified as targeted attacks~\cite{fedrecattack,pipattack,a-hum} and untargeted attacks~\cite{FedAttack}, depending on whether the objective is to promote the exposure of specific items desired by the attacker.
Targeted attacks have garnered greater research interest due to their superior stealth and greater destructive potential.
Generating effective poisonous gradients for targeted attacks presents a challenge, especially when benign users' data is made private in their personalized models within an FRS.
Existing studies resolve the challenge from various perspectives:
(1) \emph{User Embedding Approximation}. \FRA{}~\cite{fedrecattack} approximates benign users' personalized models using a small public part of their interactions;
(2) \emph{Item Popularity Enhancement}. \PIP{}~\cite{pipattack} trains a popularity estimator using items' popularity levels and enhances the target items' popularity within the estimator model, instead of manipulating scores in the personalized models of benign users.
(3) \emph{Interaction Function Poisoning}.
\ARA{} and \AHUM{}~\cite{a-hum} amplify target items' scores by modifying the learnable interaction function employed by the FRS, regardless of benign users' personalized models.
However, these attacks possess certain limitations.
\FRA{} and \PIP{} necessitate prior knowledge (i.e., benign users' public interactions in \FRA{} or items' popularity levels in \PIP{}), which is usually unavailable in practice.
\ARA{} and \AHUM{} only work in DL-FRS since the interaction function is non-learnable in MF-FRS.
Overall, we summarize the features of existing targeted attacks in~\cref{tab:comparison}{}. Different from these studies, the targeted attack proposed in this paper does not rely on prior knowledge and is not constrained by the type of FRS.

\begin{table}[!htbp]
\centering{
\caption{Comparison of targeted model poisoning attacks.}
\label{tab:comparison}
\begin{tabular}{l|ccc}
\toprule
\multirow{2}{*}{{Attack Methods}} & \multicolumn{3}{c}{{Property}}                                           \\ \cmidrule{2-4} 
& {Prior Knowledge} & {MF-FRS} & {DL-FRS}\\ 
\midrule
\FRA{} \cite{fedrecattack}       & historical interactions       &\checkmark  &\checkmark\\
\PIP{} \cite{pipattack}       & items' popularity         &\checkmark  &\checkmark     \\
\ARA{} and \AHUM{} \cite{a-hum}       & \textbf{not required}       & $\times$
& \checkmark    \\
\model{} (ours)     & \textbf{not required}                   &\checkmark  &\checkmark     \\ \bottomrule
\end{tabular}
}
\end{table}

\smallskip
\noindent\textbf{Defense Methods for FRS}. 
The defense methods against model poisoning attacks have been well studied in conventional federated learning.
These defenses focus on normalizing the uploaded gradients~\cite{NormBound} or 
filtering those gradients using statistic information such as median and trimmed mean~\cite{Media-TrimmedMean} or similarity information~\cite{Krum-MultiKrum} or the combination of the two~\cite{Bulyan}.
Their motivation is that poisonous gradients are few but unnormal compared to gradients from benign clients.
However, these defenses fail to combat targeted attacks in FRS that upload poisonous gradients to promote specific target items (cf.\ theoretical analysis in \cref{ssec:defense_analysis}).
In FRS, items are trained only on their relevant interaction data. Hence, given a target item that is usually unpopular with very sparse interaction data, poisonous gradients generated for it usually dominate and can hardly be identified via normalization or filtering.
This has been evidenced by our experiments reported in \cref{exp:defense_performance}.
To overcome the limitation, our defense method introduces two regularization terms during benign user training, restricting attack attempts based on item popularity enhancement or user embedding approximation while maintaining recommendation performance (cf.\ \cref{ssec:defense_design}).

\section{Preliminaries}
\label{ssec:preliminaries}

\cref{ssec:FR-fram} and \cref{ssec:setting-attack} present the framework of federated recommendations and the fundamental settings of the attack against them, respectively.


\subsection{Framework of Federated Recommendations} 
\label{ssec:FR-fram}

Let $u_i$ and $v_j$ be the $i$-th user and the $j$-th item in FRS, respectively.
The predicted score $\hat{x}_{ij}$ for the pair of user $u_i$ and item $v_j$ is computed as
$
\hat{x}_{ij} = \Psi(\mathbf{u}_i, \mathbf{v}_j),
$
where $\mathbf{u}_i$, $\mathbf{v}_j$, and $\Psi(\cdot)$ refers to $u_i$'s embedding, $v_j$'s embedding, and the interaction function, respectively.

The interaction function differs in MF-FRS and DL-FRS. 
Specifically, the interaction function in MF-FRS is defined as
\begin{equation*}
\Psi^\text{MF}(\mathbf{u}_i, \mathbf{v}_j) = \mathbf{u}_i \odot \mathbf{v}_j,
\end{equation*}
where $\odot$ is the dot product operator. While the interaction function in DL-FRS is defined as a stack of $L$ Multi-Layer Perceptions (MLPs):
\begin{equation}\label{equation:DL_interaction_function}
\begin{aligned}
\Psi^\text{DL}(\mathbf{u}_i, \mathbf{v}_j) & = \operatorname{sigmoid}(\mathbf{h}^\mathsf{T} \cdot \phi_L(\ldots(\phi_1(\mathbf{u}_i \oplus \mathbf{v}_j)))), \\
\phi_l(\mathbf{x}) & = \operatorname{ReLU}(\mathbf{W}_l^\mathsf{T}\mathbf{x} + \mathbf{b}_l), \,\, l = 1, \ldots, L,
\end{aligned}
\end{equation}
where $\phi_l(\mathbf{x})$ refers to the $l$-th ($1 \leq l \leq L$) layer MLP associated with learnable weights $\mathbf{W}_l$ and bias $\mathbf{b}_l$, $\mathbf{h}$ and $\oplus$ denotes the projection vector and vector concatenation operator, respectively, and $\operatorname{sigmoid}(\cdot)$ and $\operatorname{ReLU}(\cdot)$ are commonly-used activation functions.

In both MF-FRS and DL-FRS, the parameter of the personalized model of each client $u_i$ is $\{\mathbf{u}_i\}$.
For the global model, its parameter set in MF-RFS is the $m$ items' embeddings $\{\mathbf{v}_1, \ldots, \mathbf{v}_m\}$, while the counterpart in DL-FRS is 
$\{\mathbf{v}_1, \ldots, \mathbf{v}_m, \mathbf{W}_1, \ldots, \mathbf{W}_L, \mathbf{b}_1, \ldots, \mathbf{b}_L, \mathbf{h}\}$.

Let $x_{ij}$ denote the ground truth score for pair $(u_i, v_j)$ such that $x_{ij}=1$ if $u_i$ has interacted with $v_j$ and $x_{ij}=0$ otherwise.
Before training the FRS, each user $u_i$ prepares its private training dataset ${D}_i = {D}^{+}_i \cup {D}^{-}_i$, where ${D}^{+}_i=\{v_j \mid x_{ij} =1 \}$ and ${D}^{-}_i=\{ v_j \mid x_{ij}=0 \}$ contains the interacted and uninteracted items for $u_i$, respectively.
Typically, ${D}_i$ is sampled with a fixed ratio $q$ of $|{D}^{+}_i|$ to $|{D}^{-}_i|$\footnote{We set $q=1$ in the implementation, following a previous study \cite{fedrecattack}.}.
Next, the training phase takes a number of communication rounds where the $r$-th round is performed as follows.
\begin{enumerate}[leftmargin=*]
\item The server randomly selects a set ${U}^r$ of users from the whole user set and sends the current global model to them.
\item With the received global model and its local personalized model, each user $u_i \in {U}^r$ computes its loss function
\begin{equation} \label{equation:rs-loss}
    \mathcal{L}_i =-\frac{1}{|{D}_i|}\sum_{v_j \in {D}_i} x_{ij}\log\hat{x}_{ij} + (1-x_{ij})\log(1-\hat{x}_{ij}),
\end{equation}
and derives the gradients of parameters in both global and personalized models.
Following the state-of-the-art method \AHUM{}~\cite{a-hum}, the widely-used Binary Cross-Entropy (BCE) loss\footnote{Without loss of generality, our method still works for other popular loss functions. We report the empirical studies in the supplementary material~\cite{github-pieck-supple}.} is employed to quantify the difference between the model-predicted score $\hat{x}_{ij}$ and the ground truth $x_{ij}$.
\item Each user $u_i$ uploads the gradients\footnote{Gradients are represented using a symbol $\nabla$ throughout the paper.} for the global model parameters, and updates its personalized model locally:
\begin{equation*}
\mathbf{u}_i = \mathbf{u}_i - \eta\cdot\nabla\mathbf{u}_i,
\end{equation*}
where $\eta$ is a unified learning rate specified by the server.
\item Once received the gradients uploaded by all users in ${U}^r$, the server aggregates the gradients through a function $\operatorname{Agg}(\cdot)$ to update its global model.
For both MF-FRS and DL-FRS, any item embedding $\mathbf{v}_j$ is updated as
\begin{equation*}
\mathbf{v}_j = \mathbf{v}_j-\eta\cdot 
\operatorname{Agg}(\{\nabla\mathbf{v}_j^i\ |\ u_i\in{U}^r, v_j\in{D}_i\}).
\end{equation*}
Extra parameters in DL-FRS (cf.\ \cref{equation:DL_interaction_function}) are updated as
\begin{equation*}
\begin{aligned}
\mathbf{W}_l = & \mathbf{W}_l-\eta\cdot \operatorname{Agg}(\{\nabla\mathbf{W}_l^i \mid u_i\in{U}^r\}),\\
\mathbf{b}_l = & \mathbf{b}_l-\eta\cdot \operatorname{Agg}(\{\nabla\mathbf{b}_l^i \mid u_i\in{U}^r\}),\\
\mathbf{h} = & \mathbf{h}-\eta\cdot \operatorname{Agg}(\{\nabla\mathbf{h}^i \mid u_i\in{U}^r\}).
\end{aligned}
\end{equation*}
When no defense is employed, $\operatorname{Agg}(\cdot)$ performs a simple sum operation. However, a defense method can make $\operatorname{Agg}(\cdot)$ more complex.
In 
\cref{ssec:settings}, we include a series of complex defense methods for comparisons.
\end{enumerate}

Using the trained model, users can generate their own recommendation lists by selecting those uninteracted items with the top-$K$ predicted scores.

\subsection{Fundamental Settings of the Attack}
\label{ssec:setting-attack}

Let $\bar{{U}}$ and $\tilde{{U}}$ be the set of \textbf{benign users} and \textbf{malicious users} injected by the attacker, respectively.
The whole user set ${U}=\bar{{U}}\cup\tilde{{U}}$.
We proceed to introduce the attack settings from the following three aspects. 

\smallskip
\noindent\textbf{(1) Attacker Knowledge}.
To ensure the attack's applicability across various scenarios, we strictly adhere to the standard settings of FRS such that the attacker possesses only the following knowledge from the controlled malicious users:
\begin{enumerate}[leftmargin=*]
\item the attacker knows the learning rate $\eta$\footnote{Please refer to the supplementary material~\cite{github-pieck-supple} for an additional discussion on the effect of using inconsistent learning rates at the server and clients.}, a global hyperparameter for federated model training \cite{a-hum,fedrecattack,pipattack};
\item the attacker knows the structure of the base recommender model, which is shared by the server and all users;
\item the attacker knows the global model for the current $r$-th communication round only if one of the malicious users is selected for training, formally, $\tilde{{U}}\cap{U}^r\neq\mathcal{\varnothing}$.
\end{enumerate}
The aforementioned represents the essential \emph{minimum} knowledge required for a user to participate in FRS training, and also for an attacker to launch a possible attack.
Apart from that, the attacker has no access to any other information, e.g., benign users' personalized models, the gradients uploaded from benign users, items' popularity levels, and any part of the historical interactions (cf.\ \cref{tab:comparison}).

\smallskip
\noindent\textbf{(2) Attacker Capability}.
The attacker is unable to interfere with benign users, but has full control over the controlled malicious users. 
Hence, the attacker can generate well-crafted gradients and upload them via malicious users, so as to poison the FRS.
However, the attacker is unable to directly set the parameters of the global model at will because 1) the benign users among the selected users also upload gradients and 2) the aggregation function, possibly equipped with defense methods, limits the impact of poisonous gradients.

\smallskip
\noindent\textbf{(3) Attack Goal}.
Given a set ${T}$ of target items designated, 
the attacker uploads poisonous gradients during the training of FRS via controlled malicious users, aiming to increase the exposure of the target items to the benign users in the recommendation. 
This can be achieved by maximizing the average \emph{Exposure Ratio at rank $K$} (ER@$K$) for each target item $v_j \in {T}$ as
\begin{equation}
\label{equation:er}
\max \Big( 1/|{T}| \sum_{v_j \in {T}} \big(\text{ER}_{j}\text{@}K \big) \Big), \,\,\, \text{ER}_{j}\text{@}K = \frac{|\bar{{U}}_j|}{|\bar{U} \setminus \bar{{U}}'_j|},
\end{equation}
where $\bar{{U}}_j$ denotes the set of benign users{} whose top-$K$ \emph{recommendation lists} contain $v_j$ and $\bar{{U}}'_j$ denotes the set of benign users who have already \emph{interacted} with $v_j$.
However, the ER@$K$ metric in \cref{equation:er} is non-differentiable and discontinuous, thus hardly optimized.
Instead, we introduce the BCE loss (cf.\ \cref{equation:rs-loss}) as a proxy, which is defined on all pairs of a target item $v_j \in {T}$ and a benign user $u_i \in \bar{{U}}$:
\begin{equation} \label{att-loss-total}
\begin{aligned}
\tilde{\mathcal{L}}
&=-\frac{1}{|\bar{{U}}| |{T}|} \sum_{u_i\in\bar{{U}}}\sum_{v_j\in{T}} 1\cdot\log\hat{x}_{ij} + 0\cdot\log(1-\hat{x}_{ij}) \\
&= -\frac{1}{|\bar{{U}}| |{T}|} \sum_{u_i\in\bar{{U}}}\sum_{v_j\in{T}} \log \Psi(\mathbf{u}_i, \mathbf{v}_j).
\end{aligned}
\end{equation}
Minimizing $\tilde{\mathcal{L}}$ above equals to maximizing the mean predicted scores of all target items for all benign users, which inherently increases the likelihood of the items being recommended.
Optimizing this surrogate loss is an effective means to enhance the chance that benign users will include targeted items in their top-$K$ recommendations, i.e., ER@$K$.
This approach is in line with the established attack~\cite{a-hum}.
As each user embedding $\mathbf{u}_i$ is stored privately at the client's personalized model, the attacker can attempt to corrupt the FRS via the controlled malicious users' global models in two aspects:
\begin{enumerate}[leftmargin=*]
\item Each malicious user produces and uploads the poisonous gradients associated with the learnable parameters of the interaction function $\Psi(\cdot)$.
However, this approach becomes ineffective in the case of MF-FRS, where the interaction function is non-learnable.

\item Each malicious user produces and uploads the following poisonous gradient for promoting each target item $v_j \in {T}$:
\begin{equation} \label{eq:poisonous_gradients_1}
\nabla\tilde{\mathbf{v}}_j = - 1/|\bar{{U}}| \cdot \sum\nolimits_{u_i\in\bar{{U}}}\frac{\partial}{\partial \mathbf{v}_j} \Psi(\mathbf{u}_i, \mathbf{v}_j),
\end{equation}
where $\Psi(\cdot)$ can be either $\Psi^\text{MF}$ in the MF-FRS case or $\Psi^\text{DL}$ in the DL-FRS case (cf.\ \cref{ssec:FR-fram}). 
However, \cref{eq:poisonous_gradients_1} indicates that this approach requires access to embeddings of benign users, i.e., $\{ \mathbf{u}_i \mid u_i\in\bar{{U}} \}$, which have been made private in the FRS.

\end{enumerate}

We proceed to present our method \model{} (Popular Item Embedding based attaCK).
\model{} focuses on poisoning the recommender model via producing effective $\nabla\tilde{\mathbf{v}}_j$ as described in \cref{eq:poisonous_gradients_1}. Thus, \model{} is agnostic to both MF-FRS and DL-FRS.
Moreover, \model{} requires no access to benign users' embeddings. This is made possible by leveraging the unique properties of FRS to effectively identify popular items and utilizing their embeddings to generate poisonous gradients.
\section{Popular Item Embedding based Attack}
\label{sec:attack}
\begin{figure*}[ht]
\centering
    \includegraphics[width=\textwidth]{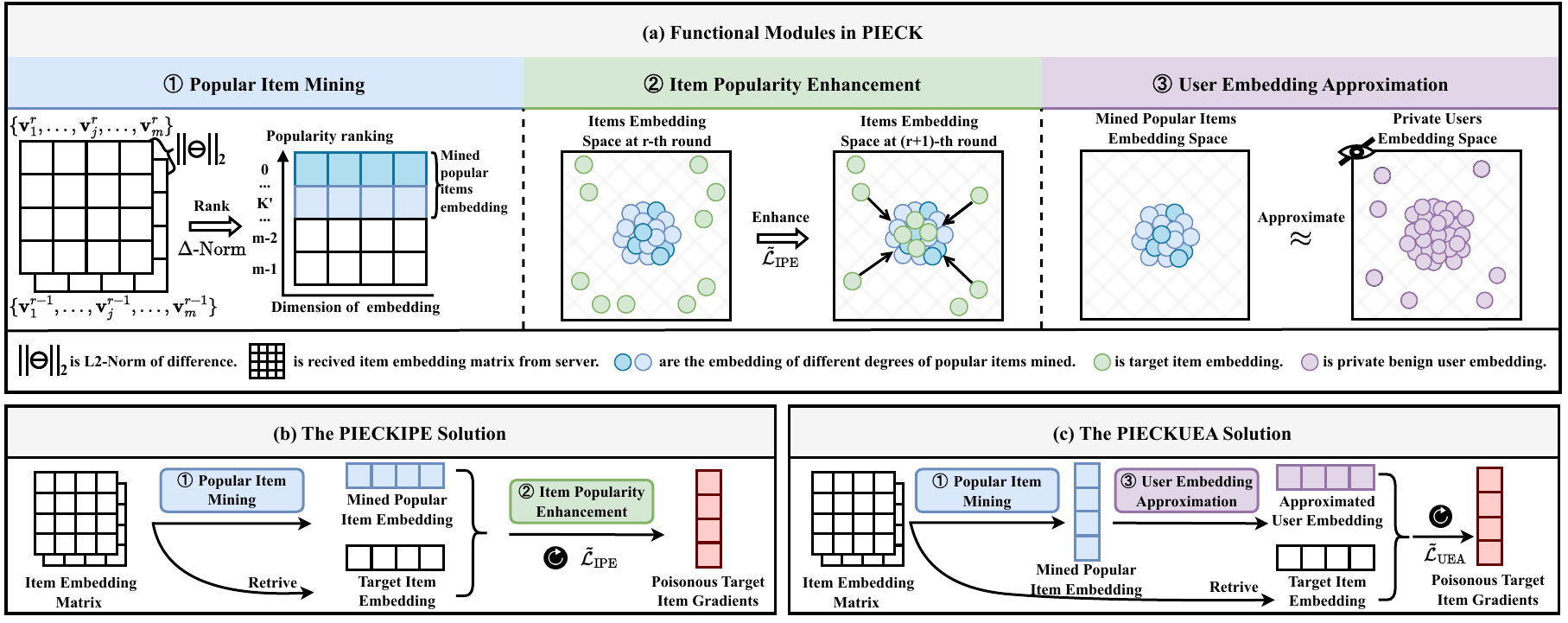} 
    \caption{Overview of \model{}: (a) functional modules in \model{}, (b) the \modelI{} solution, and (c) the \modelII{} solution.} 
    \label{fig:fram}
\end{figure*}
The overall framework of \model{} is depicted in \cref{fig:fram}.
Referring to \cref{fig:fram}(a), \model{} comprises three functional modules: \textcircled{1} popular item mining, \textcircled{2} item popularity enhancement, and \textcircled{3} user embedding approximation. Module \textcircled{1} serves as the core module and can be combined with either Module \textcircled{2} or Module \textcircled{3} to formulate two distinct attack solutions \modelI{} and \modelII{}, as shown in \cref{fig:fram}(b) and (c).
Next, we will elaborate on the basic idea of \model{} in \cref{ssec:overview} and delve into the details of the three modules in \cref{ssec:popular_items,ssec:popular_enhancement,ssec:user_embed_app}.

\subsection{Basic Idea of \model{}}
\label{ssec:overview}

Resuming the attack goal in \cref{ssec:setting-attack}, to create a model-agnostic targeted attack against FRS, we need to generate effective poisonous gradients for item embeddings without accessing benign users' embeddings.
Previous study \PIP{}~\cite{pipattack} has demonstrated that inherent popularity bias in recommender models leads to high predicted scores for popular items across the majority of users. Thus, manipulating popular items would be an effective means of generating poisonous gradients to promote target items.
However, \PIP{} assumes the attacker's prior knowledge of globally popular items, a supposition that often diverges from reality.
Getting rid of this strong assumption, we propose a \emph{popular item mining} method, which only leverages the partial knowledge that malicious users can obtain through the training of FRS in the federated setting. The method will be detailed in \cref{ssec:popular_items}.

After successfully identifying the popular items on malicious users, the attack can be initiated by generating poisonous gradients using the embeddings of identified popular items.
We explore two distinct strategies.
On the one hand, an \emph{item popularity enhancement} strategy derives poisonous gradients that align the embeddings of target items with the embeddings of those popular items, thereby increasing the exposure of target items.
This strategy and the corresponding attack solution \modelI{} will be presented in \cref{ssec:popular_enhancement}.

On the other hand, a \emph{user embedding approximation} strategy proposes to approximate the invisible embedding $\mathbf{u}_i$ of each benign user in \cref{eq:poisonous_gradients_1} to derive the poisonous gradient $\nabla\tilde{\mathbf{v}}_j$.
This strategy originates from empirical observations in typical MF-FRS \cite{ammad2019federated,FedRec,FedRec++,9162459} and DL-FRS \cite{WangYCYZZ22,perifanis2022federated,jiang2022fedncf}, where popular items' embeddings are co-trained with a significant number of users' embeddings on symmetric model structures for items and users. As a result, the distributions of popular items and user embeddings become closer.
Therefore, approximating user embeddings with popular item embeddings to derive poisonous gradients can lead to surprisingly decent performance.
The strategy along with its attack solution \modelII{} will be described in \cref{ssec:user_embed_app}.

\subsection{Popular Item Mining in FRS}
\label{ssec:popular_items}
\begin{figure}[!tbp]
\centering
     \subfigure[MovieLens-100K (ML-100K)]{
     \centering
    \begin{minipage}{0.47\linewidth}
        \label{subfig:ml-100k-dis}
        \includegraphics[width=4cm]{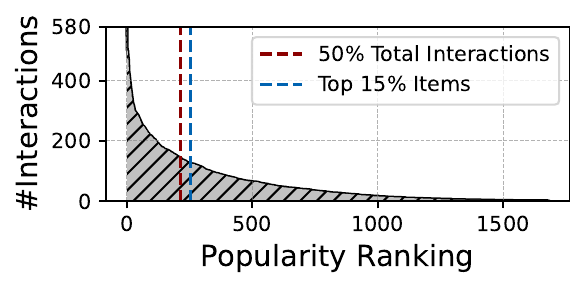}
    \end{minipage}
    }%
   \subfigure[Amazon Digital Music (AZ)]{
   \centering
    \begin{minipage}{0.47\linewidth}
        \label{subfig:az-dis}
        \includegraphics[width=4cm]{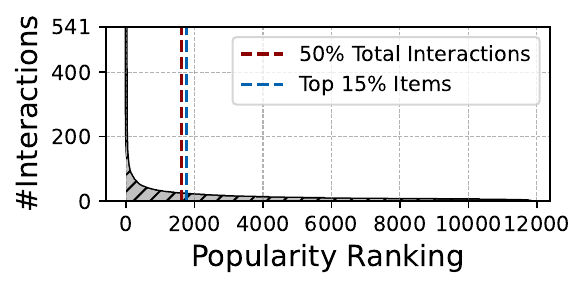}
    \end{minipage}
    }%
    \caption{The distribution of items' popularity.}
    \label{fig:inter_dis}
\end{figure}

An item's popularity is defined as the number of user interactions it receives. 
In recommendations, the popularity of items typically follows a long-tail distribution \cite{zhang2021model} --- only a small portion of items have extremely high popularity, while the rest are low.
\cref{fig:inter_dis} visualizes such long-tail phenomenon on two datasets: MovieLens-100K (ML-100K) and Amazon Digital Music (AZ). These datasets will be further described in \cref{ssec:settings}.
For clarity, we define the top $15\%$ of items with the highest number of interactions (indicated by blue dotted lines) as popular, while the remaining items are considered unpopular.
On ML-100K and AZ, the number of popular items accounts for less than one-seventh and one-sixth of the total number of items, respectively. However, interactions with these popular items exceed 50\% of the total number of interactions (indicated by red dotted lines).

Let the term $\mathcal{L}_{ij} = x_{ij}\log\hat{x}_{ij} + (1-x_{ij})\log(1-\hat{x}_{ij})$ in \cref{equation:rs-loss} be the prediction loss of an item $v_j$ to a user $u_i$. 
Thus, the overall objective of the FRS at the $r$-th round is written as
\begin{equation}
    \label{equa:global_loss}
    \mathcal{L}^r = \sum\nolimits_{v_j \in {D}_i}\big( \sum\nolimits_{u_i\in {U}_
{j}^{r}}\mathcal{L}_{ij} \big),
\end{equation}
where ${U}_{j}^{r}$ represents the set of users sampled for item $v_j$.

Let $v_j$ be a popular item and $v_k$ be an unpopular item, the following property is derived.

\begin{figure}[tbp]
\centering
     \subfigure[MF-FRS]{
     \centering
    \begin{minipage}{0.47\linewidth}
        \label{subfig:appro-pop-item-1}
        \includegraphics[width=4cm]{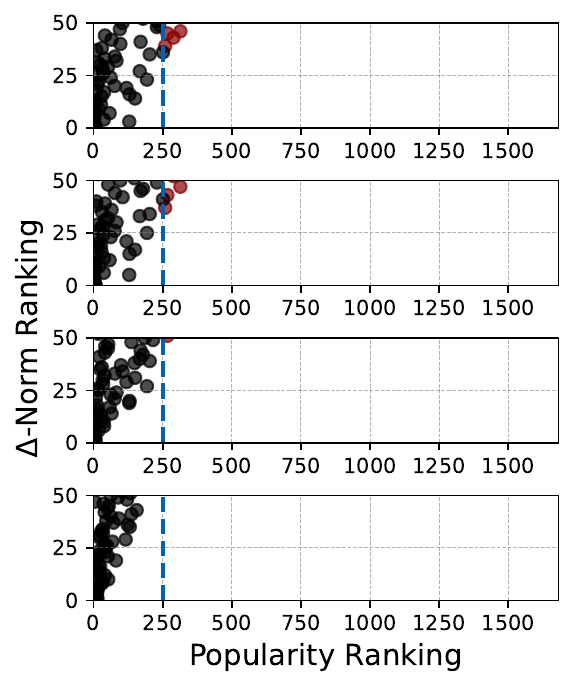}
    \end{minipage}
    }%
   \subfigure[DL-FRS]{
   \centering
    \begin{minipage}{0.47\linewidth}
        \label{subfig:appro-pop-item-2}
        \includegraphics[width=4cm]{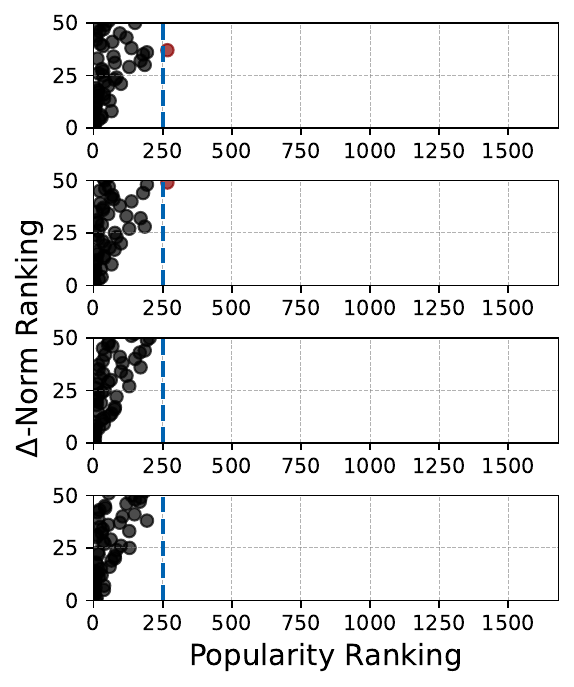}
    \end{minipage}
    }%
    \caption{Listed from top to bottom, are the popularity rankings of the top-50 items in $\Delta$-Norm for rounds 4, 8, 20, and 80.}
    \label{fig:appro-pop-item}
\end{figure}

\begin{property}\label{property_1}
\textit{
A popular item embedding $\mathbf{v}_j$ typically takes a longer time to converge compared to an unpopular item embedding $\mathbf{v}_k$.}
\end{property}

The aforementioned long-tail distribution of item popularity reveals that the user set ${U}_j^{r}$ associated with the popular item $v_j$ is typically much larger than ${U}_k^{r}$ associated with the unpopular item $v_k$.
Consequently, for the loss $\mathcal{L}^r$ in \cref{equa:global_loss}, there will be more loss terms $\mathcal{L}_{ij}$ attributed to a popular item than an unpopular one. This ultimately makes it much more prolonged for those popular items to converge.

\begin{property}\label{property_2}
\textit{
During each training round, the popular item embedding $\mathbf{v}_j$ tends to undergo larger changes compared to the unpopular item embedding $\mathbf{v}_k$.}
\end{property}

The explanation for Property~\ref{property_2} is relatively straightforward. When considering a popular item and an unpopular item in the training process, the gradients derived from \cref{equa:global_loss} for the popular item, which have been shown more difficult to converge (cf.\ Property~\ref{property_1}), will be larger. This indicates that the popular item's embedding usually experiences a greater degree of change across two training rounds.

All in all, we assert that \emph{the changes in popular items' embeddings are greater in magnitude and longer-lasting when compared to unpopular items}.
To validate this, we conduct preliminary experiments on the ML-100K dataset for both MF-FRS and DL-FRS.
To quantify the change in $v_j$'s embedding per round, we introduce the $\Delta$-Norm concept, which measures the $L_2$ Norm between the embeddings of $v_j$ at consecutive rounds. Formally, $\Delta$-Norm of item $v_j$ at the $r$-th round is 
\begin{equation}
\Delta\text{-Norm}_j^r = ||\mathbf{v}_j^{r+1}-\mathbf{v}_j^r||_2,
\end{equation}
where $\mathbf{v}_j^r$ is $v_j$'s embedding at the $r$-th round.

\cref{fig:appro-pop-item} depicts the popularity rankings of the top-50 items in $\Delta$-Norm at rounds 4, 8, 20, and 80.
Initially (rounds 4 and 8) in both MF-FRS and DL-FRS, a few unpopular items (red points to the right of the blue dotted line) enter the top-50. This is due to the fact that the embeddings of certain unpopular items are still undergoing convergence, resulting in large gradients during the early training stages.
However, as training progresses, particularly after only 20 rounds, the presence of unpopular items among the top-50 diminishes. 
Conversely, we consistently observe significant $\Delta$-Norm for popular items, which aligns with Property~\ref{property_1}.
Across all fixed rounds, the majority of the top-50 positions in $\Delta$-Norm are dominated by popular items. This observation supports our argument that the convergence of most unpopular items leads to smaller gradient updates. Meanwhile, popular items, which are still being fitted, undergo larger changes in their embeddings, consistent with our Property~\ref{property_2}.
Taken together, the sustained and prominent presence of popular items among the top-50 items in $\Delta$-Norm serves as compelling evidence to support our assertion.

The verified assertion allows us to design the popular item mining algorithm based on the $\Delta$-Norm measures, which is outlined in \cref{alg:popular_items}.
The algorithm iterates over $\tilde{R}$ times and accumulates the $\Delta$-Norm values for each item (line~1). 
The calculation of $\Delta$-Norm values, as illustrated in Module \textcircled{1} of \cref{fig:fram}(a), adopts matrix parallel computation and thus is highly efficient.
Finally, it outputs a set $P$ containing the top-$N$ items with the highest accumulated $\Delta$-Norm values. By accumulating the $\Delta$-Norm values over $\tilde{R}$ times, the algorithm ensures a stable result in identifying popular items.
In practice, we set a relatively small yet practically useful value $2$ to $\tilde{R}$ to obtain the popular items at each malicious user client.

\begin{algorithm}[!htbp]
\caption{\textsc{PopularItemMining} (mining times $\tilde{R}$, mined popular item number $N$)}
\label{alg:popular_items}
\begin{algorithmic}[1]
\While{\text{times of user being sampled }$\tilde{r} \leq \tilde{R}+1$}
    \For{each item $v_j$}
        \If{$\tilde{r} = 1$}
            $\Delta\text{-Norm}_j \gets 0$
        \Else
            ~$\Delta\text{-Norm}_j \gets \Delta\text{-Norm}_j + ||\mathbf{v}_j^{r}-\mathbf{v}_j^{r-1}||_2$
        \EndIf
    \EndFor
\EndWhile
\State \Return ${P} = \{ v_j \mid \Delta\text{-Norm}_j \text{ in top-$N$ ones} \}$
\end{algorithmic}
\end{algorithm}

\subsection{\modelI{} using Item Popularity Enhancement}
\label{ssec:popular_enhancement}

In line with our discussion in \cref{ssec:overview}, we propose an item popularity enhancement strategy to promote the recommendation exposure of the target items. This is achieved by aligning their embeddings to those of the mined popular items.
The process is depicted in Module \textcircled{2} in \cref{fig:fram}(a).

To effect this alignment, our strategy aims to optimize the {cosine similarity}\footnote{We empirically compare alternative similarity metrics in \cref{ssec:ablation_study}.} between each target item and each popular item in the embedding space. 
To aggregate these pairwise similarities effectively, we employ a \emph{weighted mean} approach, where weights assigned to each cosine similarity, are predicated on the popularity rank of the corresponding popular item in ${P}$. As such, more popular items receive higher weights.

However, the distinct characteristics of popular items (e.g., two songs of different genres) can cause the cosine similarity between the target item and popular items to lean either positively or negatively during optimization. 
This can result in overfitting towards the more prominent direction, causing a bias in the target item's embedding. 
Nonetheless, we believe that popular items in less common directions are still valuable, as they typically represent different popular feature information. Therefore, we intend to define a metric that can capture high cosine similarity between a target item and only those popular items having shared characteristics with it.

To achieve this, given a target item $v_j$, we first split the mined popular item set ${P}$ into two subsets: positive and negative popular items, based on their cosine similarity with $\mathbf{v}_j$.
They are represented as ${P}^\text{+}_j = \{\mathbf{v}_k \mid \text{cos}(\mathbf{v}_k, \mathbf{v}_j) > 0\}$ and ${P}^\text{-}_j = \{\mathbf{v}_k \mid \text{cos}(\mathbf{v}_k, \mathbf{v}_j) \leq 0\}$.
We then compute the \emph{weighted mean cosine similarity} for each subset and combine them to adjust the update magnitudes for actual popular items. This leads us to define our loss function for the attack as follows:
\begin{equation} \label{pieckipe_loss}
\begin{aligned}
\tilde{\mathcal{L}}_{\text{IPE}} = -\frac{1}{|{T}|} \sum\limits_{v_j \in {T}} \sum\limits_{\text{*}\in\{\text{+},\text{-}\}}\frac{\sum\nolimits_{v_k\in{P}^\text{*}_j} \kappa(v_k) \cdot \text{cos}(\mathbf{v}_k,\mathbf{v}_j)}{\lambda^{-1} \cdot|{P}^\text{*}_j|},
\end{aligned}
\end{equation}
where $\kappa(v_k)$ means the item $v_k$'s normalized inverse rank in ${P}_j^{*}$.
To be specific, $\kappa(v_k)$ is an alignment weight, giving more importance to the features of more popular items.
And $\lambda \in (0,1]$ regulates the strength of the weighting.
A smaller $\lambda$ intensifies the suppression of updates for the dominant direction while promoting larger updates for the rare direction.
The concept introduced in \cref{pieckipe_loss} aims to steer a target item towards greater alignment with the majority of popular items, while also considering the shared characteristics from various types of popular items. As a result, the loss defined in \cref{pieckipe_loss} comprehensively enhances the target item's popularity.

The \modelI{} solution, presented in \cref{alg:pieckipe}, is applied to each malicious user.
It starts by mining the set ${P}$ of popular items using the \textsc{PopularItemsMining} algorithm until user $u_i$ has been sampled more than $\tilde{R}$ times (line~1).
If the malicious user is subsequently sampled to participate in FRS training (line~2), it calculates the attack loss defined in \cref{pieckipe_loss} (line~3) and uploads the poisonous gradients derived from the calculated loss to manipulate the recommendation (line~4).

\begin{algorithm}
\caption{\modelI{} (mining times $\tilde{R}$, mined popular item number $N$, target item set ${T}$)}
\label{alg:pieckipe}
\begin{algorithmic}[1]
\State ${P} \gets$ \Call{PopularItemMining}{$\tilde{R},N$} \Comment{\cref{alg:popular_items}}
\While{\text{times of user being sampled }$\tilde{r} \geq \tilde{R}+1$}
    \State compute $\tilde{\mathcal{L}}_\text{IPE}$ using \cref{pieckipe_loss}
    \State upload poisonous gradients $\{ \tilde{\nabla}{\mathbf{v}}_j \mid \frac{\partial}{\partial \mathbf{v}_j} \tilde{\mathcal{L}}_\text{IPE},v_j \in {T} \}$
\EndWhile
\end{algorithmic}
\end{algorithm}

Unlike \PIP{}~\cite{pipattack}, which relies on pre-existing popular item information for enhancing popularity, \modelI{} can function independently at each client without needing this information. This is facilitated by automatically identifying popular items using the $\Delta$-Norm obtained from FRS training. However, as the recommender model becomes more personalized over time, it may prove difficult to recommend even the most popular items to all users. Consequently, in later stages of FRS training, the exposure of target items, whose popularity is enhanced by \modelI{}, may diminish (cf.\ \cref{subfig:ipe_uea_compare_mf_ml-1m}).
Thus, we provide an alternative attack solution in \cref{ssec:user_embed_app}.

\subsection{\modelII{} using User Embedding Approximation}
\label{ssec:user_embed_app}

To enhance the attack robustness throughout the FRS training, an effective way might be to directly optimize the scores of target items among all benign users based on \cref{eq:poisonous_gradients_1}.
However, this method entails knowing benign users' embeddings, typically inaccessible in FRS.
To overcome the limitation, we propose the user embedding approximation strategy by extracting useful knowledge from the mined popular items.

Let $\mathbf{V}_{P}={\{\mathbf{v}_k \mid v_k \in {P}\}}$ be the set of popular items’ embeddings.
The following property is derived.

\begin{property}\label{property_3}
\textit{
The distribution of embeddings in $\mathbf{V}_{P}$, is likely to resemble that of users, given the shared training process in the symmetric recommender model structure.}
\end{property}

This is a key observation as typical MF-FRS~\cite{ammad2019federated,FedRec,FedRec++,9162459} and DL-FRS~\cite{WangYCYZZ22,perifanis2022federated,jiang2022fedncf} are both built on user-item mutual interactions, creating symmetric recommender model structures for users and items. 
The symmetry, in turn, leads to similar embedding distributions for items and their interacting users. Such an effect is particularly pronounced for popular items that have interacted with a majority of users --- their embedding distribution closely mirrors that of all users.

To verify the observation, we conduct preliminary experiments on ML-100K, without the participation of malicious users. 
This enables us to obtain pure popularity ranking and access users' historical interactions and their embeddings during training.
Let $\mathbf{U}_{P}=\{\mathbf{u}_i \in {U} \mid \exists v_k \in {P}: x_{ik} = 1 \}$ denote the set of embeddings for users whose historical interactions include at least one item in ${P}$.
We compute the \emph{user coverage ratio} (UCR) of the popular item set ${P}$ as $|\mathbf{U}_{P}| / |{U}|$.
We then measure the similarity between the distributions of $\mathbf{V}_{P}$ and $\mathbf{U}_{P}$ using the \textit{average pairwise KL divergence} (PKL):
\begin{equation} \label{mean—pair-kld}
\text{PKL}(\mathbf{V}_{P},\mathbf{U}_{P}) = \frac{1}{|\mathbf{V}_{P}|}\frac{1}{|\mathbf{U}_{P}|}\sum\limits_{\mathbf{v}_k\in\mathbf{V}_{P}}\sum\limits_{\mathbf{u}_i\in\mathbf{U}_{P}} \text{KL}(\mathbf{v}_k, \mathbf{u}_i).
\end{equation}
A smaller PKL indicates a higher similarity between the two distributions.
Next, we collected $\mathbf{V}_{P}$ and $\mathbf{U}_{P}$ for the top-$N$ mined popular items after 200 training rounds (at which the models had converged). The PKL and UCR measures for different $N$ values are presented in \cref{tab:sim_kld}.

\begin{table}[!htbp]
\centering
\setlength{\tabcolsep}{3.2mm}
\caption{Preliminary experiments on ML-100K at 200-th round.}
\label{tab:sim_kld}
\begin{tabular}{c|c|cccc}
\toprule
\multirow{2}{*}{Metric} & \multirow{2}{*}{Model} & \multicolumn{4}{c}{Size of Popular Items Set $N$} \\ \cmidrule{3-6} 
& & 1 & 10 & 50 & 150 \\ \midrule
\multirow{2}{*}{PKL} & MF-FRS & 0.0848 & 0.0828 & 0.0871 & 0.0891 \\
& DL-FRS & 0.0026 & 0.0025 & 0.0016 & 0.0010 \\
\rowcolor[HTML]{F2F2F2} UCR & both & 0.6151 & 0.9830 & 0.9979 & 1.0 
\\ \bottomrule
\end{tabular}
\end{table}

\cref{tab:sim_kld} shows that UCR rapidly reaches 0.9830 at $N$ = 10, indicating that mined popular items cover most users swiftly. 
In DL-FRS, PKL is consistently small and unaffected by $N$ variations.
PKL in MF-FRS initially reduces, then rises with a peak similarity between $\mathbf{V}_{P}$ and $\mathbf{U}_{P}$ distributions at $N$ = 10.
In general, selecting $N$ as 10 or more obtains desired values for PKL and UCR in both MF-FRS and DL-FRS. This highlights the notable alignment in the distribution patterns of popular items and users during the FRS training process.

Motivated by these findings, we revise the original attack loss in \cref{att-loss-total}, replacing the set of inaccessible embeddings of benign users (i.e., $\{ \mathbf{u}_i \mid u_i\in\bar{{U}} \}$) with the embeddings of mined top-$N$ popular items (i.e., $\{ \mathbf{v}_k \mid v_k\in{P} \}$): 
\begin{equation} \label{pieckuea-loss}
\tilde{\mathcal{L}}_\text{UEA} = -\frac{1}{N} \frac{1}{|{T}|} \sum\nolimits_{v_k\in{P}}\sum\nolimits_{v_j\in{T}} \log\Psi (\mathbf{v}_k, \mathbf{v}_j).
\end{equation}

As an alternative attack solution executed at each malicious client, \modelII{} is presented in \cref{alg:pieckuea}.
The preparation of mining popular items (line~1) follows the same process as described in \cref{alg:pieckipe}. However, the poisonous gradients are calculated using \cref{pieckuea-loss} and then uploaded to the server in each participation round (lines~2--4). Notably, each embedding $\mathbf{v}_k$ used for approximation is treated as a constant in this process and is excluded in the backpropagation.

\begin{algorithm}
\caption{\modelII{} (mining times $\tilde{R}$, mined popular item number $N$, target item set ${T}$)}
\label{alg:pieckuea}
\begin{algorithmic}[1]
\State ${P} \gets$ \Call{PopularItemMining}{$\tilde{R},N$} \Comment{\cref{alg:popular_items}}
\While{$\text{times of user being sampled } \tilde{r} \geq \tilde{R}+1$}
    \State compute $\tilde{\mathcal{L}}_\text{UEA}$ using \cref{pieckuea-loss}
    \State upload poisonous gradients $\{ \tilde{\nabla}{\mathbf{v}}_j \mid \frac{\partial}{\partial \mathbf{v}_j} \tilde{\mathcal{L}}_\text{UEA}, v_j \in {T} \}$
\EndWhile
\end{algorithmic}
\end{algorithm}

\section{Defense Analysis and New Design}
\label{sec:defense}

\cref{sec:attack} has presented our identified attack \model{}, notable for its applicability to different base model types without requiring any prior knowledge.
Despite its potential to cause substantial harm to FRS applications, this threat has largely been overlooked.
In \cref{ssec:defense_analysis}, we will analyze the present defense methods in the federated setting and explain their inadequacies to combat targeted model poisoning in FRS.
In \cref{ssec:defense_design}, we will devise an innovative defense method aimed at reducing security risks associated with \model{} and other targeted model poisoning in FRS.

\subsection{Defense Analysis}
\label{ssec:defense_analysis}

Denote the proportion of malicious users by $\tilde{p} = |\tilde{{U}}| / |{U}|$.
As aforementioned, at each communication round $r$, the server randomly selects a batch of users ${U}^r$ to participate in the training.
In conventional federated learning, for each learnable parameter, selected benign users upload the normal gradient, while selected malicious users upload the poisonous version.
The expected proportion of poisonous gradients received by the server is $\tilde{E} = \frac{|{U}^r|\cdot\tilde{p}}{|{U}^r|}=\tilde{p}$.

However, in FRS, a benign user uploads the normal gradient of an item embedding $\mathbf{v}_j$ only if that item $v_j$ is in their private training datasets.
The expected number of such users at a round is $|{U}^r|\cdot (1-\tilde{p})\cdot p_j$, where $p_j$ denotes the probability that $v_j$ is contained within a benign user's private training dataset.
Considering this factor, the expected proportion of poisonous gradients of $\mathbf{v}_j$ received by the server is:
\begin{equation}
\label{eq:exp}
\begin{aligned}
\tilde{E}(v_j)
&=\frac{|{U}^r|\cdot\tilde{p}}{|{U}^r|\cdot (1-\tilde{p})\cdot p_j + |{U}^r|\cdot\tilde{p}},\\
&=\frac{\tilde{p}}{(1-\tilde{p})\cdot p_j+\tilde{p}} \,\,\, ,
\end{aligned}
\end{equation}
where for $p_j=1$ (all selected benign users contain $v_j$ in their training datasets), $\tilde{E}$ reaches its minimum value of $\tilde{p}$.
Indeed, the probability $p_j$ is given by
\begin{equation}\label{equ:p_j}
p_j=\frac{1}{|\bar{{U}}|} \sum\nolimits_{u_i\in\bar{{U}}}p_{{ij}},
\end{equation}
where $p_{{ij}}$ is the probability that the private training dataset ${D}_i$ of a benign user $u_i \in \bar{{U}}$ contains $v_j$.

Let $|{D}^{+}_i|$ and $|{D}^{-}_i|$ be the disjoint parts of ${D}_i$ containing interacted and uninteracted items for $u_i$, respectively.
We have
\begin{equation}
p_{{ij}}=\begin{cases}
\frac{|{D}_i^-|}{|{V}|-|{D}_i^+|} \Rightarrow \frac{q \cdot |{D}_i^+|}{|{V}|-|{D}_i^+|}, & \text{if~}v_j\notin{D}_i^+,\\
1, &\text{otherwise},
\end{cases}
\end{equation}
where ${V}$ denotes the whole item set and $q$ is the sampling ratio of  $|{D}^{+}_i|$ to $|{D}^{-}_i|$ (cf.\ \cref{ssec:FR-fram}).
Considering the small ratio $q$ (typically not exceeding $4$)\footnote{
Per the findings from previous studies \cite{NCF, fedrecattack, pipattack, a-hum}, a $q$ value of 4 or less is set to avoid performance drops in the RS. See the supplementary material~\cite{github-pieck-supple} for the evaluation of the varying $q$ on a real-world dataset.}
and the fact that the number of a user's interacted items is usually significantly smaller than the total number of items (i.e., $|{D}_i^+| \ll |{V}|$), the value of $p_{{ij}}$ will be very small for the case that $v_j \notin {D}_i^+$.

Moreover, each target item $v_j \in {T}$ specified by the attacker is usually an extremely cold item. Hence, for most benign users, $v_j$ is not in ${D}_i^+$, making $p_j$ in \cref{equ:p_j} a small value.
Returning to \cref{eq:exp}, we observe that $\tilde{E}(v_j)$ increases when $p_j$ decreases. Therefore, the small value of $p_j$ for the target item $v_j$ results in $\tilde{E}(v_j)$ being significantly larger than $\tilde{p}$.

The fact that $\tilde{E}(v_j) \gg \tilde{p}$ is a critical challenge for existing defense methods\cite{NormBound,Media-TrimmedMean, Krum-MultiKrum, Bulyan} designed for conventional federated learning.
These methods rely on gradient normalization or filtering, assuming that the number of poisonous gradients must be lower than that of benign gradients.
For instance, consider the \MEDIAN{} approach \cite{Media-TrimmedMean}, which computes the median of received gradients for each dimension as the final normalized gradient and theoretically requires $\tilde{E}(v_j) < 0.5$.
However, if we assume $\tilde{E}(v_j) < 0.5$ according to \cref{eq:exp}, we deduce that $p_j > \frac{\tilde{p}}{1-\tilde{p}}$. Substituting the robust upper bound of the \MEDIAN{} approach (i.e., $\tilde{p}=0.5$) into the equation, we obtain $p_j > 1$. This contradicts the nature of any item in FRS, where $p_j \ll 1$, making the \MEDIAN{} approach ineffective.
Likewise, other methods do not escape this problem.

\subsection{New Design of the Defense Method}
\label{ssec:defense_design}

Our new design focuses on the following three key findings about the identified attack method \model{}:
\begin{enumerate}[leftmargin=*]
\item[F1] The long-tail item popularity leads to great and long-lasting changes in popular items' embeddings (cf.\ \cref{ssec:popular_items});
\item[F2] The inherent popularity bias causes most users to predict higher scores for popular items (cf.\ \cref{ssec:popular_enhancement});
\item[F3] The recommender symmetry makes close embedding distributions between popular items and users (cf.\ \cref{ssec:user_embed_app}).
\end{enumerate}

The finding F1 suggests setting a threshold to limit the magnitude of item embedding changes at consecutive rounds to combat the popular item mining from malicious clients.
However, two issues arise: (1) The difficulty of determining an appropriate threshold: A high threshold would significantly degrade the recommender system's performance, while a low threshold would fail to provide an effective defense.
(2) The method's invalidation in scenarios where item popularity has been publicly available: Malicious users can access popular items without even a mining process.

To address these issues, we introduce two additional regularization terms into the benign user training, which utilizes the findings F2 and F3. 
Regarding F2, the regularization term in \cref{regul1} is designed to induce confusion between the features of popular and unpopular items. This makes it challenging for malicious users to precisely capture the distinctive features of popular items and consequently, prevents them from counterfeiting target items as popular ones.
As for F3, the regularization term in \cref{regul2} aims to create significant separation in the embedding distribution of popular items and users. By doing so, it ensures that the user embeddings inferred from popular item embeddings are inherently inaccurate.

In our defense method, each benign user $u_i \in \bar{{U}}$ goes through the following steps:
\begin{enumerate}[leftmargin=*]

\item Mine popular item set ${P}_i$ by calling \cref{alg:popular_items};

\item Obtain a set of unpopular items $\Delta D_i = D_i \setminus P_i$.

\item Compute regularization term $\mathit{Re}_1$ as weighted mean pairwise cosine similarity between embeddings of $\Delta D_i$ and $P_i$:
\begin{equation} \label{regul1}
\mathit{Re}_1 = \frac{1}{|\Delta {D}_i|} \sum\limits_{v_j \in \Delta {D}_i} \sum_{v_k\in{P}_{i}} \kappa'(v_k) \cdot \text{cos}(\mathbf{v}_k,\mathbf{v}_j),
\end{equation}
where $\kappa'(v_k)$ is the normalized \emph{exponential}\footnote{We employed the exponential form to stipulate the term focus even more on the most popular items for improved defense effectiveness.} inverse rank of item $v_k$ in $P_i$.
Introducing $\mathit{Re}_1$ in the defense method aims to prevent item popularity enhancement in \modelI{}.

\item Compute regularization term $\mathit{Re}_2$ as average KL divergence between the distributions of user embedding and mined popular items' embeddings:
\begin{equation} \label{regul2}
\mathit{Re}_{2} =
 \sum\nolimits_{v_k\in{P}_{i}} \kappa'(v_k) \cdot \text{KL}(\mathbf{v}_k, \mathbf{u}_i).
\end{equation}
Introducing $\mathit{Re}_{2}$ aims to invalidate the user embedding approximation in \modelII{}. 

\item Participate in training using the combined defense loss:
\begin{equation} \label{defense-loss}
    \mathcal{L}^\text{def}_i = \mathcal{L}_i - \beta \cdot \mathit{Re}_{1} - \gamma \cdot \mathit{Re}_{2},
\end{equation}
where $\mathcal{L}_i$ is the original loss defined in \cref{equation:rs-loss}, and $\beta$ and $\gamma$ are non-negative trade-off parameters for the two regularization terms.
\end{enumerate}

\section{Experiments}
\label{sec:experiments}
To verify our proposals, we conduct extensive experiments on both MF-FRS and DL-FRS, involving three real datasets, four attack methods, and six defense methods.
The entire codebase and instruction on hyperparameter tuning are all made available on Github~\cite{github-pieck}. 

\subsection{Experiment Settings}
\label{ssec:settings}
\noindent \textbf{Data and Models}. We use three datasets for evaluations: {MovieLens-100K (ML-100K)} \cite{MovieLens}, {MovieLens-1M (ML-1M)} \cite{MovieLens}, and {Amazon Digital Music (AZ)} \cite{AZ}. 
To confirm the model-agnostic nature of \model{}, we adopt Matrix Factorization \cite{BPR} for MF-FRS and Neural Collaborative Filtering (NCF) \cite{NCF} for DL-FRS as the underlying base models.

\noindent \textbf{Baselines}.
Our model, \model{}, is benchmarked against leading model poisoning attacks such as \FRA{} \cite{fedrecattack}, \PIP{} \cite{pipattack}, \ARA{} \cite{a-hum}, and \AHUM{} \cite{a-hum}. To test \model{} and compare our proposed defense method, we apply established defense methods \NB{} \cite{NormBound}, \MEDIAN{} \cite{Media-TrimmedMean}, \TMEAN{} \cite{Media-TrimmedMean}, \KRUM{} \cite{Krum-MultiKrum}, \MKRUM{} \cite{Krum-MultiKrum} and \BULYAN{} \cite{Bulyan}.

\noindent
\textbf{Metrics}. 
To assess the attack effectiveness in promoting target items, we utilize the \emph{Exposure Ratio at rank $K$} (ER@$K$) as defined in \cref{equation:er}. 
For fairness,  we follow \FRA{} \cite{fedrecattack} and randomly select target items to ${T}$ from the set of uninteracted items.
To measure the recommendation performance, we employ the \emph{Hit Ratio at rank $K$} (HR@$K$) following the NCF approach \cite{NCF}. 
More comprehensive details on our setups can be found in the supplementary material~\cite{github-pieck-supple}.

\begin{table*}[tbp]
\centering{
\caption{Comparison of all attack methods on the FRS with no defense (default malicious user ratio $\tilde{p}=5\%$).}
\label{tab:overall-attack}
\renewcommand{\arraystretch}{0.12}
\setlength{\tabcolsep}{1.8mm}
\begin{tabular}{c|cc|cc|cc|cc|cc|cc}
\toprule
 & \multicolumn{6}{c|}{MF-FRS} & \multicolumn{6}{c}{DL-FRS} \\
\cmidrule{2-7} \cmidrule{8-13}
Attacks & \multicolumn{2}{c|}{ML-100K} & \multicolumn{2}{c|}{ML-1M} & \multicolumn{2}{c|}{AZ} & \multicolumn{2}{c|}{ML-100K} & \multicolumn{2}{c|}{ML-1M} & \multicolumn{2}{c}{AZ} \\
\cmidrule{2-3} \cmidrule{4-5} \cmidrule{6-7} \cmidrule{8-9} \cmidrule{10-11} \cmidrule{12-13}
 &ER@10 &  HR@10 & ER@10 & HR@10 & ER@10 &  HR@10 & ER@10 &  HR@10 & ER@10 &  HR@10 & ER@10 & HR@10 \\
\midrule
\rowcolor[HTML]{F2F2F2} \textsc{NoAttack} &0.23  &57.16 &0.00   &61.32 &0.09  &24.25 &0.00    &45.6 & 0.00 &  47.70 & 0.00 & 17.80 \\
\FRA{}                                    &0.23  &\underline{57.58} &0.00   &\textbf{61.37} &0.05  &\textbf{24.28}  &0.00   &\underline{45.28} &0.00 &\textbf{47.52} &0.00 &\underline{17.84}  \\
\PIP{}                                    &26.42 &56.95 &16.75  &61.04 &45.05 &\underline{24.24} &\textbf{100.00} &44.96 &\textbf{100.00} &  46.46 &\textbf{100.00} &17.81 \\
\ARA{}                                    &0.11  &56.95 &0.00   &61.18 &0.01  &24.18 &\textbf{100.00} &45.07 &\textbf{100.00} &  47.20 &\textbf{100.00} &17.80\\
\AHUM{}                                   &31.09 &57.05 &12.88  &61.11 &27.26 &24.05 &\textbf{100.00}  &45.07 &\textbf{100.00} &  \underline{47.33} &\textbf{100.00} & 17.77 \\
\modelI{} \textbf{(ours)} & \underline{87.47}  &\textbf{57.69} &\underline{90.47}  &\underline{61.21} &\underline{50.21}  &23.93 & \textbf{100.00}  &\textbf{45.39} & \textbf{100.00}  &  47.28 &\textbf{100.00}  & 17.80 \\
\modelII{} \textbf{(ours)} & \textbf{93.39} &\textbf{57.69} &\textbf{98.04} &61.03 &\textbf{72.11} &24.09 &\textbf{100.00} &  \textbf{45.39} &\textbf{100.00}  &  47.30 & \textbf{100.00} & \textbf{17.95} \\
\bottomrule
\end{tabular}}
\end{table*}

\begin{table*}[ht]
\centering{
\caption{Comparison of all defense methods combating the attacks on ML-100K (default malicious user ratio $\tilde{p}=5\%$).}
\label{tab:overall-defense}
\renewcommand{\arraystretch}{0.12}
\setlength{\tabcolsep}{1.8mm}
\begin{tabular}{c|cc|cc|cc|cc|cc|cc}
\toprule
& \multicolumn{6}{c|}{MF-FRS} & \multicolumn{6}{c}{DL-FRS} \\
\cmidrule{2-7} \cmidrule{8-13}
Defenses & \multicolumn{2}{c|}{\AHUM{}} & \multicolumn{2}{c|}{\modelI{}} & \multicolumn{2}{c|}{\modelII{}} & \multicolumn{2}{c|}{\AHUM{}} & \multicolumn{2}{c|}{\modelI{}} & \multicolumn{2}{c}{\modelII{}} \\
\cmidrule{2-3} \cmidrule{4-5} \cmidrule{6-7} \cmidrule{8-9} \cmidrule{10-11} \cmidrule{12-13}
& ER@10 & \multicolumn{1}{c|}{HR@10} & ER@10 & \multicolumn{1}{c|}{HR@10} & ER@10 & \multicolumn{1}{c|}{HR@10} & ER@10 & \multicolumn{1}{c|}{HR@10} & ER@10 & \multicolumn{1}{c|}{HR@10} & ER@10 & HR@10 \\
\midrule
\rowcolor[HTML]{F2F2F2} \NODEF    &31.09	&57.05	&87.47	&57.69	&93.39	&57.69 &100.00	&45.07	&100.00	&45.39	&100.00	&45.39\\
\NB{}     &\textbf{0.00}	&55.14	&\textbf{0.00}	&55.14	&\uline{0.11}	&55.25 &100.00	&\underline{45.17}	&100.00	&\textbf{45.60}	&100.00	&45.17\\
\MEDIAN{} &\textbf{0.00}	&54.83	&\textbf{0.00}	&54.72	&\textbf{0.00}	&54.93 &100.00	&43.90	&100.00	&44.86	&100.00	&44.11\\   
\TMEAN{}  &34.51*	&55.46	&81.78	&\textbf{56.73}	&48.52	&\textbf{55.89} &100.00	&\textbf{45.39}	&100.00	&45.28	&100.00	&\textbf{45.71}\\
\KRUM{}   &\textbf{0.00}	&51.22	&\textbf{0.00}	&51.22	&\textbf{0.00}	&51.33 &100.00	&43.58	&100.00	&43.16	&100.00	&44.11\\  
\MKRUM{}  &\underline{29.84}	&\uline{55.89}	&79.04	&55.89	&59.23	&\underline{55.67} &100.00	&44.96	&100.00	&\underline{45.49}	&100.00	&\underline{45.36}\\
\BULYAN{} &33.71*	&54.61	&49.43	&54.72	&79.04	&55.25 &100.00	&44.33	&100.00	&44.54	&100.00	&44.64\\ 
\textbf{ours} &\textbf{0.00} &\textbf{58.11} &\underline{1.25} &\underline{56.31} &\textbf{0.00} &\textbf{55.89} &\textbf{0.00} &43.58 &\textbf{0.00} &44.43 &\textbf{0.00} &43.69  \\
\bottomrule
\end{tabular}}
\end{table*}

\subsection{Attack Performance}
We compare all considered attack methods on three datasets, two model types, and multiple baselines, with a default proportion $\tilde{p}$ of 5\% malicious users (the effect of varying $\tilde{p}$ is studied in \cref{exp:parameters_analysis}) and report the results in \cref{tab:overall-attack}.

\subsubsection{Attack Effectiveness}
\label{ssec:attack_effectiveness_comparison}
Overall, the ER@10 measures in \cref{tab:overall-attack} indicate that our proposed attacks (\modelI{} and \modelII{}) achieve highly competitive scores compared to all baselines across both model types and three datasets. 
The slightly lower measure on the AZ dataset is attributed to its larger item pool 
which makes it more challenging for the target item to be recommended; the results also show that most attacks are effective to ML-100K, probably because its user-item interactions are relatively condensed.

The inferior performance of \PIP{} and \FRA{} stems from their lack of prior knowledge about user interaction history and item popularity levels. This limitation hinders their ability to effectively enhance item popularity and approximate user embeddings, resulting in a failure to achieve the desired increase in the target item's exposure.
Besides, \ARA{} and \AHUM{} show significantly lower effectiveness in MF-FRS compared to DL-FRS. Their attacks are specifically designed to poison the learnable interaction parameters in DL-FRS, which renders them ineffective in MF-FRS where the interaction function is fixed as a dot product.
In contrast, our approach focuses on poisoning publicly available item embeddings common to both types of models, allowing us to achieve model-agnostic attacks.
Considering that \model{} applies to both FRS types and does not rely on prior knowledge, its remarkably high attack effectiveness raises concerns about its potential to cause significant harm to practical FRS.


Upon a detailed examination of the two \model{} variants, \modelII{} demonstrates relatively better performance. To understand the reasons behind this, we provide the complete convergence curves of \modelI{} and \modelII{} for MF-FRS on ML-1M in \cref{subfig:ipe_uea_compare_mf_ml-1m}. As observed, \modelI{} experiences a more significant decline in ER@10 compared to \modelII{} as the rounds increase. We attribute this behavior to the reduced effectiveness of popularity enhancement in \modelI{} as FRS converges towards personalized user preferences. In contrast, \modelII{} directly approximates the distribution of user embeddings through mined popular items, which elevates target item scores among all users, resulting in a more robust and effective attack.
However, this comes at the expense of higher computational overhead (cf.\ cost analysis in \cref{ssec:cost_analysis}).


The \textsc{NoAttack} method unexpectedly shows small ER@10 scores in two cases (0.23\% on ML-100K and 0.09\% on AZ). This is because our target items were randomly selected to ensure fair comparisons (cf. 
\cref{ssec:settings}
) and are thus always recommendable.
Additionally, most ER@10 measures for DL-FRS are as high as 100\%, likely due to the susceptibility of its learnable interaction function to poisoning attacks.

\subsubsection{Recommendation Performance}
The HR@10 measures in \cref{tab:overall-attack} show that the recommendation performance on MF-FRS and DL-FRS remains largely unaffected by all attacks, including our proposed ones, across all three datasets. For instance, on ML-1M, the best-performing \modelII{} enhances its ER@10 by 98.04\% compared to the \textsc{NoAttack} scenario. Remarkably, the corresponding HR@10 experiences only a negligible decrease of 0.29\%.
Please note that malicious user behaviors injected by various attack methods introduce inherent randomness in FRS training, resulting in variations of HR@10 measures among the recommender models associated with different attacks. As a result, the HR@10 measures of \textsc{NoAttack} may not always be the highest.
However, it is essential to emphasize that these variations are almost negligible and do not affect our empirical finding that all attacks do not lead to a degradation of recommendation performance.

\subsection{Defense Performance}
\label{exp:defense_performance}
We evaluate the proposed defense method using the top-3 most effective attacks: \modelI{}, \modelII{}, and \AHUM{}. The evaluation primarily focuses on the ML-100K dataset, as these attacks demonstrate superior effectiveness on this dataset (cf. \cref{tab:overall-attack}). Similar trends are observed in other datasets, so their results are omitted. The results are listed in \cref{tab:overall-defense}.

\subsubsection{Attack Effectiveness}
The ER@10 measures in~\cref{tab:overall-defense} show that our proposed defense achieves the best score for both MF-FRS and DL-FRS compared to all competitor methods.
When applying our defense method, all attacks are ineffective on both types of models.
For example, compare to \textsc{NoDenfense}, for \modelII{}, we reduce ER@10 from 93.39\% to 0.00\% on MF-FRS and from 100\% to 0.00\% on DL-FRS, respectively.
Our defenses against \AHUM{} and \modelI{} show consistent performance, verifying the effectiveness of the proposed method.
Even though our defense does not include a specific regularization term for \AHUM{} explicitly, its employed combined loss function (cf.\ \cref{defense-loss}) incorporates both item embeddings and the interaction function. 
We contend that the \AHUM{} attack also estimates user embedding distributions by iteratively refining random initial user embeddings to identify those who rated the target item poorly. The $Re_2$ loss term is designed to specifically counteract this try of \AHUM{}. To summarize, our defense has proven universally effective, demonstrated by its success against three variants of FRS poisoning attacks, as outlined in \cref{fig:attacker}.

Surprisingly, we observed that certain defense methods even lead to higher ER@10 measures, compared to \textsc{NoDenfense}, as marked with * in~\cref{tab:overall-defense}.
We propose that it happens because these defenses indiscriminately normalize and filter the gradients of each item.
For non-target items, the disposal of the gradient might diminish the user's score, hence inadvertently accentuating the target item.
For the target item, the benign gradient in the minority might be filtered out, hence amplifying the impact of the poisonous gradient.
Differently, our defense will not reduce the score of non-target items and filter out the benign gradient of the target item. 

\subsubsection{Recommendation Performance}
In \cref{tab:overall-defense}, our defense demonstrates only a minor drop in HR@10 measures for both types of FRS against the three attacks. This success stems from the trade-off between the original training loss and the added regularization terms in FRS training. Such an optimized, combined defense loss in \cref{defense-loss} effectively counters potential attacks from \modelI{} and \modelII{}, while still ensuring model accuracy.

\begin{figure}[tbp]
\centering
     \includegraphics[width=1.0\linewidth]{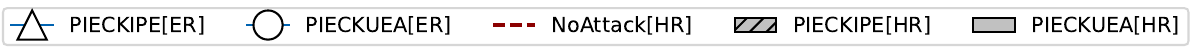}%
     \vspace{-0.2cm}
     \subfigure[Attacks vs $\tilde{p}$.]{
     \centering
    \begin{minipage}{0.48\linewidth}
        \label{subfig:attack_p_mf_ml-100k}
        \includegraphics[width=\linewidth]{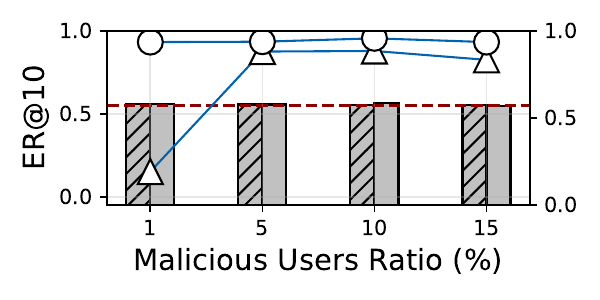}
    \end{minipage}
    }%
     \subfigure[Defense vs $\tilde{p}$.]{
     \centering
    \begin{minipage}{0.48\linewidth}
        \label{subfig:defense_p_mf_ml-100k}
        \includegraphics[width=\linewidth]{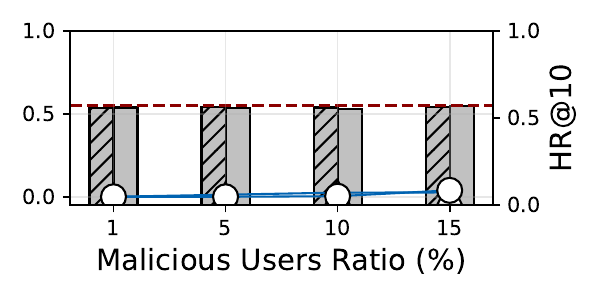}
    \end{minipage}
    }%
    \vspace{-0.2cm}
   \subfigure[Attacks vs $N$.]{
   \centering
    \begin{minipage}{0.48\linewidth}
        \label{subfig:attack_n_mf_ml-100k}
        \includegraphics[width=\linewidth]{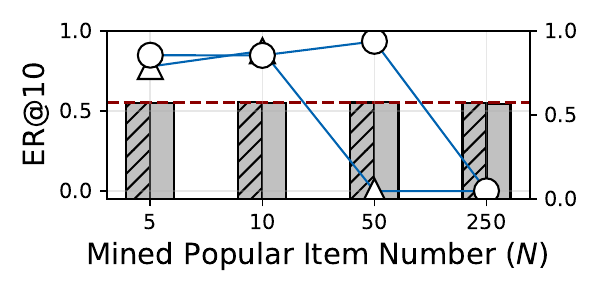}
    \end{minipage}
    }%
   \subfigure[Defense vs $N$.]{
   \centering
    \begin{minipage}{0.48\linewidth}
        \label{subfig:defense_n_mf_ml-100k}
        \includegraphics[width=\linewidth]{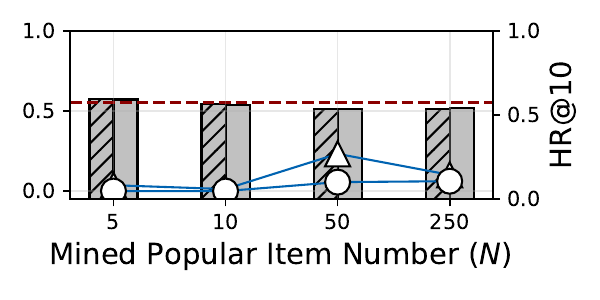}
    \end{minipage}
    }%
    \caption{Effect of $\tilde{p}$ and $N$ on proposed attacks and defense. 
    }
    \label{fig:effect_of_malicious}
\end{figure}
\subsection{Parameters Analysis}
\label{exp:parameters_analysis}
We analyze the hyperparameters $\tilde{p}$, $N$, and $K$ in MF-FRS. MF-FRS is shown in \cref{tab:overall-attack} to be more robust against attacks compared to DL-FRS, making it suitable for discerning the characteristics of various attack and defense methods.

\subsubsection{Effect of malicious users' ratio \texorpdfstring{$\tilde{p}$}{Lg}}
Referring to the attack performance shown in \cref{subfig:attack_p_mf_ml-100k}, we observe that increasing $\tilde{p}$ results in a gradual rise in the attack effectiveness (ER@10) of \modelI{} and \modelII{}, while the recommendation performance (HR@10) remains relatively stable. Notably, \modelII{} outperforms \modelI{} by directly improving the target item's score on all benign users through user embedding approximation, achieving a more robust attack compared to using item popularity enhancement in \modelI{}.
 
In terms of defense performance (reported in \cref{subfig:defense_p_mf_ml-100k}), our method significantly reduces the harm caused by \modelI{} and \modelII{} under different $\tilde{p}$ values while maintaining recommendation performance comparable to the \textsc{NoAttack} scenario. This confirms the high effectiveness of our defense against both versions of \model{}.

\subsubsection{Effect of mined popular item number \texorpdfstring{$N$}{Lg}}
Regarding the attack performance (shown in \cref{subfig:attack_n_mf_ml-100k}), an $N$ value up to 10 yields high attack effectiveness (ER@10) for \modelI{} and \modelII{}, with minimal impact on recommendation performance. However, beyond certain thresholds, specifically, $N=50$ for \modelI{} and $N=250$ for \modelII{}, the attack's effectiveness is significantly affected. This occurs because, at these points, the mined popular items may also include some unpopular ones, hindering the capture of popular item features and user embedding approximation. In conclusion, the results suggest that an appropriate number of mined popular items can effectively enhance the attack strategies employed in both \modelI{} and \modelII{}.

Regarding the defense performance (shown in \cref{subfig:defense_n_mf_ml-100k}), a moderate number of mined popular items ($N=10$) effectively reduces the attack effectiveness while maintaining decent recommendation performance. However, using a large $N$ not only reduces defense effectiveness but also decreases recommendation performance. This happens because the defense method with a large $N$ may focus on constraining relatively unpopular items, which are not targeted by malicious users but instead negatively impact recommendation performance.

\subsubsection{Effect of the number \texorpdfstring{$K$}{Lg} of recommended items}
Referring to \cref{tab:attack_k}, when $K$ varies from 5 to 20, both variants of \model{} demonstrate effective attacks when our defense is not deployed (\modelI{}/\modelII{} vs \textsc{NoDefense}). However, once the defense is deployed, it successfully counters \model{} under different values of $K$ (\modelI{}/\modelII{} vs \textbf{ours}). Additionally, both our proposed attack and defense have minimal impact on recommendation performance (HR@$K$). These results validate the robustness and insensitivity of our attacks and defense to changes in $K$.

\begin{table}[tb]
\centering{
\caption{Effect of $K$ on proposed attacks and defense.}
\label{tab:attack_k}
\renewcommand{\arraystretch}{0.7}
\setlength{\tabcolsep}{2mm}
\begin{tabular}{c|c|cc|cc}
\toprule
\multirow{2}{*}{Attacks} & \multirow{2}{*}{Defenses}& \multicolumn{4}{c}{Performance (MF-FRS and ML-100K)}\\
\cmidrule{3-6} &
 &ER@5 &  HR@5 & ER@20 &  HR@20  \\
\midrule
\rowcolor[HTML]{F2F2F2} \textsc{NoAttack}  &\textsc{NoDefense} &0.00  &38.92  &0.46	  &76.78\\
\modelI{}          &\textsc{NoDefense} &\underline{85.99} &\underline{39.34}  &\underline{90.21}  &\underline{76.88}\\
\modelI{}          &\textbf{ours}      &0.46  &\underline{39.34}  &6.95	  &\textbf{77.52}\\
\modelII{}         &\textsc{NoDefense} &\textbf{93.62} &38.71  &\textbf{93.51}  &\textbf{77.52}\\
\modelII{}         &\textbf{ours}      &0.00  &\textbf{39.66}  &0.68	  &75.93\\
\bottomrule
\end{tabular}}
\end{table}

\subsection{Ablation Study} \label{ssec:ablation_study}

First, we examine the efficacy of techniques used in $\mathcal{L}_\text{IPE}$ (cf.\ \cref{pieckipe_loss} for \modelI{}), namely the average pairwise cosine similarity (PCOS for short) metric, the weighted term $\kappa(\cdot)$, and the subset partitioning policy (${P}^\text{+/-}$). We do not explore \modelII{} in this context since it does not involve complex techniques like those in \modelI{}.

Referring to the results in \cref{tab:abla-combined} (left), we find that the PCOS metric outperforms another widely-used metric PKL (cf.\ \cref{mean—pair-kld}) in terms of attack effectiveness, attributing to its better capture of latent features in their respective embeddings. 
The use of the weighted term ($\kappa(\cdot)$) further improves attack effectiveness by assigning larger weights to learn from relatively more popular items. 
Finally, employing the subset partitioning policy (${P}^\text{+/-}$) results in \modelI{} achieving decent performance, emphasizing the necessity of capturing shared features among different types of popular items.

Next, we analyze the efficacy of two regularization terms ($\mathit{Re}_{1}$ and $\mathit{Re}_{2}$) in the defense loss $\mathcal{L}^\text{def}$ in \cref{defense-loss}. 
The ER@10 measures in \cref{tab:abla-combined} (right) indicate that using both regularization terms jointly yields favorable defense performance. This highlights the importance of finely tuning item and user embeddings while concurrently constraining item popularity enhancement and user embedding approximation for effective defense.
In contrast, using only one regularization term either harms recommendation performance ($\mathit{Re}_1$ only) or fails to provide sufficient defense effectiveness ($\mathit{Re}_2$ only). 

\begin{table}[tb]
    \centering
    \caption{Ablations of $\mathcal{L}_\text{IPE}$ \& $\mathcal{L}^\text{def}$ of MF-FRS on ML-100K.}
    \renewcommand{\arraystretch}{0.7}
    \setlength{\tabcolsep}{0.32mm}
    \begin{tabular}{c|cc|cc||cc|cc|cc}
        \toprule
        \multicolumn{3}{c|}{$\mathcal{L}_\text{IPE}$} &\multicolumn{2}{c||}{$\modelI{}$} & \multicolumn{2}{c|}{$\mathcal{L}^\text{def}$}  & \multicolumn{2}{c|}{\modelI{}} & \multicolumn{2}{c}{\modelII{}} \\
        \midrule
        Metric  & $\kappa(\cdot)$ & ${P}^\text{+/-}$ & ER@10 &HR@10 & $\mathit{Re}_1$ & $\mathit{Re}_2$ & ER@10 &HR@10 & ER@10 &HR@10 \\
        \midrule
        \rowcolor[HTML]{F2F2F2} PKL & & &39.75 &57.48 & & &87.47 &57.69 &93.39 &57.69\\
        & & &46.01 &56.95  &+ & &\underline{26.2} &55.46 &\textbf{0.00} &54.19 \\
        PCOS&+ & &\underline{71.41} &\underline{57.58} & &+  &49.2 &\textbf{56.63} &\underline{69.36} &\textbf{56.73}\\
        &+ &+ &\textbf{87.47} &\textbf{57.69} &+ &+ 
    &\textbf{1.25} &\underline{56.31} &\textbf{0.00} &\underline{55.89} \\
        \bottomrule
    \end{tabular}
    \label{tab:abla-combined}
\end{table}

\begin{figure}[tbp]
\centering
     \subfigure[Performance trends on MF-FRS.]{
     \centering
    \begin{minipage}{0.48\linewidth}
        \label{subfig:ipe_uea_compare_mf_ml-1m}
        \includegraphics[width=\linewidth]{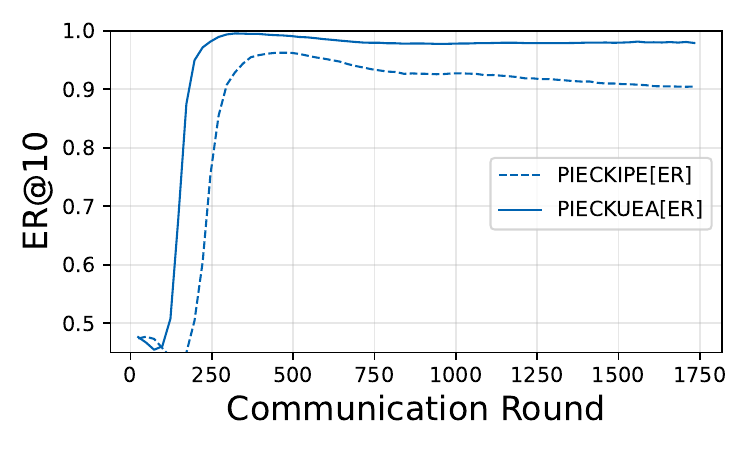}
    \end{minipage}
    }%
     \subfigure[Cost analysis on MF/DL-FRS.]{
     \centering
    \begin{minipage}{0.48\linewidth}
        \label{subfig:cost_mf_dl_ml-1m}
        \includegraphics[width=\linewidth]{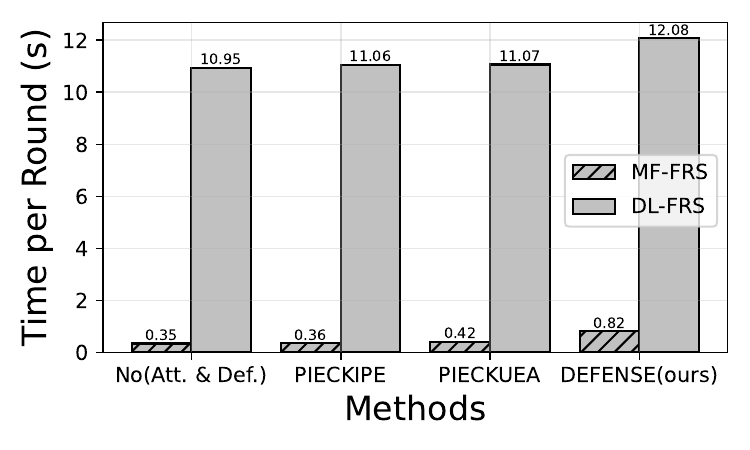}
    \end{minipage}
    }%
    \caption{Performance trends and cost analysis on ML-1M.}
    \label{fig:trends_cost}
\end{figure}

\subsection{Cost Analysis}
\label{ssec:cost_analysis}
We conduct 500 rounds of evaluation for the proposed attacks and defense and report the average time cost per training round in \cref{subfig:cost_mf_dl_ml-1m}. DL-FRS requires more time compared to MF-FRS due to the usage of a learnable interaction function.
Regarding the attack, we observe a negligible time increase for both 
MF-FRS and DL-FRS compared to the vanilla scenario (denoted as No(Att.\& Def.)). 
This is because the three \model{} modules can be efficiently executed via matrix parallel.
\cref{subfig:cost_mf_dl_ml-1m} also shows that \modelII{} exhibits a bit higher time cost than \modelI{}. 
This discrepancy arises from \modelII{} requiring mining more popular items, and implementing multiple rounds in batches (default batch size is 5 and round size is 3) to improve the score of the target item on approximated user embeddings.
For defense, the average time increase per round for MF-FRS and DL-FRS is only 0.47 and 1.13 seconds, respectively, compared to No(Att.\& Def.).
These slight increments are acceptable for maintaining FRS security. Overall, our attacks and defense are efficient and effective.

\subsection{Discussion on System Settings}
\label{ssec:system_setting_discussion}

\begin{table}[tb]
\centering{
\caption{Effect of $q$ and $|T|$ on proposed attacks and defense.}
\label{tab:other_effects}
\renewcommand{\arraystretch}{0.5}
\setlength{\tabcolsep}{1.8mm}
\begin{tabular}{c|c|cc|cc}
\toprule
\multirow{2}{*}{Attacks} & \multirow{2}{*}{Defenses} & \multicolumn{2}{c|}{$q=10$} &\multicolumn{2}{c}{$|T|=3$}\\
\cmidrule{3-6} &
 &ER@10 &  HR@10 & ER@10 &  HR@10  \\
\midrule
\rowcolor[HTML]{F2F2F2} \textsc{NoAttack}  &\textsc{NoDefense}     &0.00	            &57.26                &0.23  &57.16         \\
\modelI{}                                  &\textsc{NoDefense}     &\underline{30.75}   &\underline{57.58}    &\underline{59.16} &57.26    \\
\modelI{}                                  &\textbf{ours}          &1.03	            &55.99                &0.21  &57.58           \\
\modelII{}                                 &\textsc{NoDefense}     &\textbf{89.86}      &\textbf{57.79}       &\textbf{93.93} &\underline{57.90}  \\
\modelII{}                                 &\textbf{ours}          &0.68	            &54.51                &0.50  &\textbf{58.43}             \\
\bottomrule
\end{tabular}}
\end{table}

\subsubsection{Large sample ratio}
We scrutinize how our attack and defense approaches perform when the FRS uses an extremely large sample ratio $q$ under MF-FRS and ML-100K.
Empirical evidence, as reported in the supplementary material~\cite{github-pieck-supple}, justifies setting $q=10$ since FRS performance markedly declines beyond this point.
The results in~\cref{tab:other_effects} confirm that both the proposed attack ($N$=$10$ for \modelI{} and $N$=$15$ for \modelII{}) and defense methods sustain effectiveness at this elevated $q$ level without diminishing FRS performance.
The ER@10 of \modelI{} is suboptimal, which we attribute to the user's training bias towards uninteracted items, resulting in gradient updates of popular items (most are interacted items) being diluted. This, however, does not impact \modelII{}.




\subsubsection{Multiple target items}

Multi-target scenarios might be interesting in practice. In the supplementary material~\cite{github-pieck-supple}, we explore how varying $|T|$ values and different attack strategies affect our proposals.
The findings suggest that training just one target item and uploading $|T|$ copies to the server is quite effective.
This technique alleviates the interference of multiple target item updates and makes optimization simple. Moreover, it involves no extra training or a vast number of malicious clients.
As a result, we have employed this simple and cost-effective strategy to assess our attack and defense methods.
\cref{tab:other_effects} reports the system performance on the MF-FRS+ML-100K scenario with $|T|=3$ and malicious client user fixed to 5\%.
As can be seen, our attack achieves a high ER@$10$ even with multiple targets, and our defense remains very strong when compared to the default configuration of $|T|=1$.
For primary experiments, we set $|T|=1$, aiming for easy controls of parameters such as $N$, $\tilde{p}$, $\beta$, $\gamma$ in evaluations, which also aligns with existing attack methods~\cite{a-hum,fedrecattack,pipattack}.

\section{Conclusion and Future Work}

This study introduces \model{}, a model-agnostic and prior-knowledge-free targeted poisoning attack for practical federated recommender systems (FRS). 
We propose \modelI{} and \modelII{} as two diverse solutions to increase the exposure of target items based on effectively mining popular items during FRS training. Existing federated defenses have been found ineffective against \model{}, leading us to propose a new defense method with two well-designed regularization terms. Extensive experiments across model types, datasets, attacks, and defenses have validated the efficacy of our proposals.

For future work, it is interesting to explore collaborative defense methods that combine both server-side and client-side strategies. 
It is also interesting to extend our attack and defense methods to content-based federated recommendations.

\section*{Acknowledgment}
This work is supported by the National Key R\&D Program of China (No.~2022YFB3304100), the Pioneer R\&D Program of Zhejiang (No.~2024C01021), and the Major Research Program of Zhejiang Provincial Natural Science Foundation (No.~LD24F020015).


\clearpage
\balance
\bibliographystyle{plain}
\bibliography{sample-base.bib}


\clearpage
\section*{\large Supplementary Material to the Paper Entitled ``Preventing the Popular Item Embedding Based Attack in Federated Recommendations''}

\bigskip

\subsection{Details of the Experiment Settings}
\label{ssec:settings}
\subsubsection{Datasets} 
\label{sssec:datasets}

We use three datasets for evaluation, namely {MovieLens-100K (ML-100K)} \cite{MovieLens}, {MovieLens-1M (ML-1M)} \cite{MovieLens}, and {Amazon Digital Music (AZ)} \cite{AZ}.
The statistics are shown in \cref{tab:data_sta}, where AZ features a lower interaction rate compared to the other two while all datasets from real-world scenarios exhibit high sparsity.
Such sparsity typically makes it challenging for FRS to accurately learn users' preferences and achieve effective recommendations.
Following a previous study \cite{he16fast}, for each user, we adopt the leave-one-out method to create the training and test sets.

\bigskip
\begin{table}[!htbp]
    \centering{
    \caption{Dataset statistics (`Rate' equals $\text{\#(Interactions)} / $ $\text{\#(Users)}$ and `Sparsity' equals $1-\text{\#(Interactions)} / (\text{\#(Users)} \times \text{\#(Items)})$).}
    \label{tab:data_sta}
    \begin{tabular}{c|ccccc}
    \toprule
    {Dataset} & {\#(Users)}  & {\#(Items)}  & {\#(Interactions)} & {Rate} & {Sparsity} \\ \midrule
    ML-100K  & 943    & 1,682  & 100,000      & 106  & 93.70\%  \\
    ML-1M    & 6,040  & 3,706  & 1,000,209    & 166  & 95.53\%  \\
    AZ       & 16,566 & 11,797 & 169,781      & 10   & 99.91\%  \\ \bottomrule
    \end{tabular}
    }
\end{table}
\smallskip

\subsubsection{System Settings}

To confirm the model-agnostic nature of \model{}, we adopt Matrix Factorization \cite{BPR} for MF-FRS and Neural Collaborative Filtering (NCF) \cite{NCF} for DL-FRS as the underlying base models. Despite the existence of various dedicated recommender models with intricate components, the above two models have gained widespread adoption in practice and have been extended to the federated setting~\cite{fedrecattack,pipattack,a-hum}. Their appeal lies in their ability to ensure high generalization and low communication overhead, making them highly representative choices for our evaluation.
In our setting, each user in the dataset is regarded as a `client' in the federation. 
The batch size of randomly selected users per round for MF-FRS is 256 on ML-100K and ML-1M datasets and 1024 on AZ. For DL-FRS, the batch size remains 256 for all datasets.
We train MF-FRS and DL-FRS with learning rates $\eta=1.0$ and $\eta=0.005$, where corresponding model parameters are set the same as previous studies \cite{fedrecattack,a-hum}.
Other parameter settings, related to the popular item mining algorithm and the defense loss can be found in our instruction~\cite{github-pieck}.

\subsubsection{Attack Baselines}

We compare \model{} with the following state-of-the-art model poisoning attacks:

\begin{itemize}[leftmargin=*]

\item \FRA{} \cite{fedrecattack} accesses a \emph{portion} of benign users' historical interactions to approximate their embeddings.

\item \PIP{} \cite{pipattack} involves explicit promotion and popularity enhancement for target items.

\item \ARA{} \cite{a-hum} randomly initializes users' embeddings at malicious users, specifically designed for DL-FRS.

\item \AHUM{} \cite{a-hum} extends \ARA{} by enhancing the attack based on mining hard users for increased effectiveness.

\end{itemize}

For a fair comparison without utilizing prior knowledge, we mask the historical interactions for \FRA{} and popularity levels of items for \PIP{} in the implementation. Moreover, we set null parameters for \ARA{} when applying it to MF-FRS. 

\subsubsection{Defense Methods}
\label{sssec:defense_methods}

To evaluate \model{} and compare our proposed defense method, we apply different defense methods to the aggregation function $\operatorname{Agg}(\cdot)$ on the server and tune them \emph{optimal}. They process the uploaded gradients as follows.

\begin{itemize}[leftmargin=*]

\item \NB{} \cite{NormBound}: A thresholding approach is employed to bound the $L_2$ Norm of all gradients uploaded by users.

\item \MEDIAN{} \cite{Media-TrimmedMean}: The median of received gradients for each dimension is computed.

\item \TMEAN{} \cite{Media-TrimmedMean}: The $\tilde{p}$ largest and smallest values for each dimension are removed, and the rest are averaged.

\item \KRUM{} \cite{Krum-MultiKrum}: The most similar gradient from received gradients in the squared Euclidean norm space is selected. 

\item \MKRUM{} \cite{Krum-MultiKrum}: The ($2\tilde{p}$) least similar gradients produced by \KRUM{} are removed iteratively with the rest averaged.

\item \BULYAN{} \cite{Bulyan}: Gradients are selected by \MKRUM{} and then averaged using \TMEAN{}.

\end{itemize}

\subsubsection{Evaluation Metrics} \label{ssec:evaluation_metrics}
In attacking and defending an FRS, both the \textbf{effectiveness of attack} and the \textbf{recommendation performance} are crucial considerations. An effective attack should successfully promote target items without significantly degrading the performance of the system to avoid detection.

To assess the attack effectiveness in promoting target items, we utilize the \emph{Exposure Ratio at rank $K$} (ER@$K$) as defined in \cref{equation:er}. Attacks like \PIP{} and \AHUM{} specifically focus on the least popular items, which allows for more significant promotion opportunities since the server receives fewer benign gradients. However, for fairness, we follow \FRA{} \cite{fedrecattack} and randomly select target items to ${T}$ from the set of uninteracted items.
To measure the recommendation performance, we employ the \emph{Hit Ratio at rank $K$} (HR@$K$) following the NCF approach \cite{NCF}. A higher HR@$K$ indicates more accurate recommendations of the FRS.

\subsection{Effect of Sampling Ratio $q$}

\begin{figure}[!htbp]
\centering
    \label{subfig:ipe_uea_compare_mf_ml-1m}
    \includegraphics[width=\linewidth]{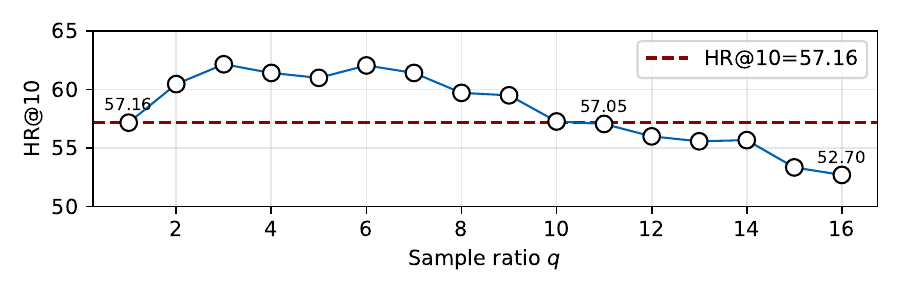}
    \caption{The effect of sample ratio $q$ of MF-FRS on ML-100K.}
    \label{fig:effect_of_q}
\end{figure}
\smallskip

\begin{table*}[tb]
\centering{
\caption{Effect of $|T|$ on proposed attacks and defense on MF-FRS and ML-100K.}
\label{tab:effect_of_T}
\renewcommand{\arraystretch}{0.4}
\setlength{\tabcolsep}{1.4mm}
\begin{tabular}{c|c|cc|cc|cc|cc|cc|cc}
\toprule
   & & \multicolumn{8}{c|}{\texttt{Train Together}}    & \multicolumn{4}{c}{\texttt{Train One Then Copy}} \\
\cmidrule{3-14}
Attacks & Defenses & \multicolumn{2}{c|}{$|T|=2$} &\multicolumn{2}{c|}{$|T|=3$} &\multicolumn{2}{c|}{$|T|=4$} &\multicolumn{2}{c|}{$|T|=5$} &\multicolumn{2}{c|}{$|T|=3$} &\multicolumn{2}{c}{$|T|=5$}\\
\cmidrule{3-14} &
 &ER@10 &  HR@10 & ER@10 &  HR@10 & ER@10 &  HR@10 & ER@10 &  HR@10  & ER@10 &  HR@10 & ER@10 &  HR@10 \\
\midrule
\rowcolor[HTML]{F2F2F2} \textsc{NoAttack}  &\textsc{NoDefense}     &0.00	            &57.26               &0.23  &57.16            &0.23  &57.16             &0.23  &57.16 &0.23  &57.16 &0.23  &57.16                                   \\
\modelI{}                                  &\textsc{NoDefense}     &\underline{75.61}   &\textbf{57.58}   &\underline{57.09} &57.48    &\underline{39.55} &57.16             &\underline{32.70} &\underline{56.84}  &\underline{59.16} &57.26  &\underline{54.76} &56.73             \\
\modelI{}                                  &\textbf{ours}          &0.80	            &57.48               &0.46  &\textbf{58.01}   &0.05 &\underline{57.90}               &0.02  &\textbf{58.22}  &0.21  &57.58 &2.33  &\underline{58.01}                              \\
\modelII{}                                 &\textsc{NoDefense}     &\textbf{92.72}      &\underline{57.05}      &\textbf{87.33} &57.05   &\textbf{85.24} &56.73              &\textbf{81.84} &56.31 &\textbf{93.93} &\underline{57.90} &\textbf{97.65} &57.69                      \\
\modelII{}                                 &\textbf{ours}          &0.00	            &56.2 &0.00         &57.69       &0.00        &\textbf{58.11}        &0.13  &\textbf{58.22} &0.50  &\textbf{58.43}  &0.34  &\textbf{58.32}                                      \\
\bottomrule
\end{tabular}}
\end{table*}

\begin{table}[tb]
\centering{
\caption{Effect of inconsistent learning rates on our attacks.}
\label{tab:abla_learningrate}
\renewcommand{\arraystretch}{0.4}
\setlength{\tabcolsep}{1.5mm}
\begin{tabular}{c|cc|cc|cc}
\toprule
\multirow{2}{*}{$\eta_i$} & \multicolumn{2}{c|}{\textsc{NoAttack}}& \multicolumn{2}{c|}{\modelI{}} & \multicolumn{2}{c}{\modelII{}}\\
\cmidrule{2-7} 
 &ER@10 &  HR@10  &ER@10 &  HR@10 & ER@10 &  HR@10  \\
\midrule
\rowcolor[HTML]{F2F2F2} 1e-0  &0.23  &57.16 &87.47     &57.69  &93.39	  &57.69\\
1e-2                          &0.00   &49.31 &98.06     &48.99  &95.56  &48.36\\
1e-2$\sim$1e-0                      &0.00 &21.74 &55.13 &22.16  &55.69  &21.95\\
\bottomrule
\end{tabular}}
\end{table}

\begin{table}[htb]
\centering{
\caption{Effect of loss function on proposed attacks and defense.}
\label{tab:abla_loss}
\renewcommand{\arraystretch}{0.4}
\setlength{\tabcolsep}{1.9mm}
\begin{tabular}{c|c|cc|cc}
\toprule
\multirow{2}{*}{Attacks} & \multirow{2}{*}{Defenses}& \multicolumn{2}{c|}{BCE} & \multicolumn{2}{c}{BPR}\\
\cmidrule{3-6} &
 &ER@10 &  HR@10 & ER@10 &  HR@10  \\
\midrule
\rowcolor[HTML]{F2F2F2} \textsc{NoAttack}  &\textsc{NoDefense} &0.23              &57.16             &0.00	            &57.26\\
\modelI{}                                  &\textsc{NoDefense} &\underline{87.47} &\textbf{57.69}    &\underline{83.14} &\textbf{57.48}\\
\modelI{}                                  &\textbf{ours}      &1.25              &56.31             &4.67	            &54.08\\
\modelII{}                                 &\textsc{NoDefense} &\textbf{93.39}    &\underline{57.69} &\textbf{90.21}    &\underline{56.95}\\
\modelII{}                                 &\textbf{ours}      &0.00              &55.89             &0.00	            &53.45\\
\bottomrule
\end{tabular}}
\end{table}

Here, we discuss the effect of varying the sampling ratio $q$ on the recommendation performance.
A case study is conducted on the MF-FRS using the ML-100K dataset, in line with the primary experiments.
Our analysis reveals that the sampling ratio, $q$, critically influences the recommendation performance, with HR@$10$ peaking at intermediate values of $q$. 
When $q$ is increased from its lowest value, we observe an improvement in recommendation quality. This improvement suggests that a slight increase in $q$ enhances the algorithm's ability to predict by providing a more comprehensive capture of user-item interactions.
However, this upward trend in performance plateaus and subsequently inverts beyond a critical threshold of $q$, Specifically, when $q$ is equal to or exceeds 11, there is a marked deterioration in HR@$10$.
It is noteworthy that the decline in recommendation quality at high $q$ values is not just suboptimal; it actually falls below the quality at the lowest sampling ratio tested (i.e., $q=1$, where HR@$10$ equals 57.16\%). 
This degradation at elevated $q$ values renders the system less appealing to attackers, as the utility of the recommendations is compromised to the extent that the value derived from potential manipulation is outweighed by the intrinsic inefficacy of the system.

In conclusion, our research highlights the importance of finely tuning the sampling ratio $q$ to an optimal, smaller value. This balance is crucial to avoid overfitting while maintaining representativeness, ensuring that the system can provide high-quality recommendations for real-world usage.

\subsection{Effect of Target Item Number $|T|$}

It is interesting to know how the attacks as well as the defense perform when the attacker wants to promote multiple item numbers.
Therefore, we vary the number of target items $|T|$ from 2 to 5 and evaluate the system performance on an MF-FRS using the ML-100K dataset.
We maintain a constant proportion $\tilde{p}$ of malicious clients at 5\% to ensure a fair comparison.
The results are presented in \cref{tab:effect_of_T}, covering two different training strategies in the attack.

We first consider a \texttt{Train Together} strategy, where multiple malicious clients are involved ($N=5$ for \modelI{} and $N=50$ for \modelII{} for optimal performance) and produce poisonous gradients to promote the set of $T$ of target items jointly.
For this strategy, the results in \cref{tab:effect_of_T} shows that our attack achieves relatively high ER@$K$ and our defense is very robust.
However, as the $|T|$ increases, the efficacy of the attack diminishes. This decline is attributed to potential interference in updates between different target items, posing challenges for malicious users attempting to optimize multiple poisonous gradients simultaneously.
To counteract this, we experimented with segregating the attackers into groups with each focusing on a single item. This required an increase in the percentage of attackers to preserve attack strength, which is less efficient.

Another feasible strategy is to train only one target item while uploading $|T|$ copies of its poisonous gradient to the server, referred to as \texttt{Train One Then Copy}. This is a more straightforward and cost-efficient strategy as it eliminates the need for extensive training or a large group of attackers.
Experiments were conducted with $|T|=3, 5$, and with slight adjustments to $N$ for optimal performance ($N=15$ for \modelI{} and $N=250$ for \modelII{}), while other settings remained consistent with the \texttt{Train Together} strategy. 
The corresponding columns in \cref{tab:effect_of_T} illustrate that the approach significantly enhances the attack's effectiveness, e.g., the ER@$10$ for $|T|=5$ increasing from 81.84\% to 97.65\%.

Therefore, we decided to apply the \texttt{Train One Then Copy} strategy for the primary experiment in \cref{ssec:system_setting_discussion}.

\subsection{Discussion on Inconsistent Learning Rates in FRS}

We explore the impact of learning rate inconsistency between client and server on proposed attacks using the ML-100K dataset on the MF-FRS. Specifically, we test two different scenarios: 
1) clients using a static learning rate $\eta_i=1e-2$ unlike the server's ($\eta=1e-0$); 
2) clients with dynamic $\eta_i$ values ($1e-2$ to $1e-0$). Their results are reported in rows~2--3 of \cref{tab:abla_learningrate}.
Compared to our standard scenario (Row 1), these two scenarios incur a significant drop in recommendation quality (HR@$10$), showing the drawbacks of mismatched learning rates. 
For example, under `NoAttack', the HR@$10$ decreases by 7.85\% for the fixed inconsistency scenario (Row 2), while the dynamic inconsistency (Row 3) dropped even further by 35.42\%.
In contrast, the attack effectiveness (ER@$10$) of \textsc{Pieck*} remains robust except in the dynamic rate scenario (Row 3) where performance notably decreased (HR@$10$ fell to 21.74\% and ER@$10$ of \textsc{Pieck*} is about 55\%).
These results confirm the robustness of our attack in a stable FRS environment (no poorly configured learning rates).
Therefore, as long as the recommendation system is well-functioning, its popularity bias will exist (without our defense employed). By leveraging the aggregation operation of FRS, malicious users can exploit the three properties of popular items (cf.\ \cref{sec:attack}) based on the gradient changes, thereby achieving effective \textsc{Pieck} to poison the FRS model.

\subsection{Discussion on Other Loss Functions in FRS} 

We further discuss the generalizability of our proposed attacks and defenses to other popular loss functions.
We let the client utilize BPR loss \cite{BPR} to train MF-FRS based on ML-100K and test the performance of attack and defense.
As shown in \Cref{tab:abla_loss}, our attacks and defenses remain effective under the BPR loss, highlighting the  universality of our attack and defense strategies across different loss functions.
This universality can be attributed to the underlying optimization objectives shared by various loss functions in collaborative filtering-based recommendation systems.
Regardless of whether it is BPR, BCE loss, or other prevalent loss functions, they all fundamentally strive to prompt the recommendation system to allocate higher scores to items with user interactions and lower scores to those without.

Popular items, being accessed by a larger user, offer a distinct gradient change, allowing malicious users to distinguish them from unpopular ones within the FRS.
In a similar fashion, our customized defense strategy effectively combats these attacks.

\end{document}